\documentclass[peerreview,conference]{IEEEtran}
\IEEEoverridecommandlockouts
\usepackage{graphicx}
\usepackage{subcaption}
\usepackage{url}
\usepackage{amsfonts}
\usepackage{float}
\usepackage{amsthm}
\usepackage{listings}
\newtheoremstyle{newline_def}
    {9pt} 
    {9pt} 
    {} 
    {} 
    {\bfseries} 
    {} 
    {\newline} 
    {} 
\theoremstyle{newline_def}
\newtheorem{definition}{Definition}

\usepackage{algorithmic}
\usepackage[algo2e,ruled,vlined]{algorithm2e}
\SetCommentSty{scriptsize}
\SetKwComment{tcc}{\{}{\}}
\SetKwIF{If}{ElseIf}{Else}{if}{}{else if}{else}{endif}
\SetKwFor{While}{while}{}{endw}
\SetKwFor{ForEach}{for each}{}{endfch}

\SetKwRepeat{Do}{do}{while}

\usepackage{cite}
\usepackage{amsmath,amssymb,amsfonts}
\usepackage{algorithmic}
\usepackage{graphicx}
\usepackage{textcomp}
\usepackage{xcolor}
\def\BibTeX{{\rm B\kern-.05em{\sc i\kern-.025em b}\kern-.08em
    T\kern-.1667em\lower.7ex\hbox{E}\kern-.125emX}}
\begin{document}

\title{A More Scalable Sparse Dynamic Data Exchange}

\author{\IEEEauthorblockN{1\textsuperscript{st} Andrew Geyko}
\IEEEauthorblockA{\textit{Department of Computer Science} \\
\textit{University of New Mexico}\\
Albuquerque, USA \\
ageyko@unm.edu}\\
\and
\IEEEauthorblockN{2\textsuperscript{nd} Gerald Collom}
\IEEEauthorblockA{\textit{Department of Computer Science} \\
\textit{University of New Mexico}\\
Albuquerque, USA \\
geraldc@unm.edu}\\
\and
\IEEEauthorblockN{3\textsuperscript{rd} Derek Schafer}
\IEEEauthorblockA{\textit{Department of Computer Science} \\
\textit{University of New Mexico}\\
Albuquerque, USA \\
dschafer1@unm.edu}\\
\and
\IEEEauthorblockN{4\textsuperscript{th} Patrick Bridges}
\IEEEauthorblockA{\textit{Department of Computer Science} \\
\textit{University of New Mexico}\\
Albuquerque, USA \\
patrickb@unm.edu}\\
\and
\IEEEauthorblockN{5\textsuperscript{th} Amanda Bienz}
\IEEEauthorblockA{\textit{Department of Computer Science} \\
\textit{University of New Mexico}\\
Albuquerque, USA \\
bienz@unm.edu}\\
}

\maketitle
\IEEEpeerreviewmaketitle

\begin{abstract}
Parallel architectures are continually increasing in performance and scale, while underlying algorithmic infrastructure often fail to take full advantage of available compute power. Within the context of MPI, irregular communication patterns create bottlenecks in parallel applications.  One common bottleneck is the sparse dynamic data exchange, often required when forming communication patterns within applications.  There are a large variety of approaches for these dynamic exchanges, with optimizations implemented directly in parallel applications.  This paper proposes a novel API within an MPI extension library, allowing for applications to utilize the variety of provided optimizations for sparse dynamic data exchange methods.  Further, the paper presents novel locality-aware sparse dynamic data exchange algorithms.  Finally, performance results show significant speedups up to 20x with the novel locality-aware algorithms.
\end{abstract}


\section{Introduction}~\label{sec:intro}
Parallel architectures are continuously improving in scale, computational power, and efficiency, allowing for increasingly large and efficient high performance computing (HPC) applications.  Despite improvements in hardware capabilities, parallel applications fail to fully utilize emerging hardware due to high costs associated with data movement and inter-process communication.  Standard communication algorithms treat all communication as equal.  However, the costs of various paths of communication on current and emerging computers vary greatly with relative locations of the sending and receiving processes, with notable differences between intra-socket, inter-socket, and inter-node communication.  

A large number of parallel applications, including sparse solvers and simulations, are bottlenecked by the performance of irregular communication.  This type of communication consists of each process sending messages to a subset of other processes, with the pattern of communication dependent on the specific problem being solved such as the sparsity pattern of a given matrix.  One bottleneck in irregular communication is forming a communication pattern, determining the set of processes to which each process must send as well as the set from which it receives.  Often, such as in the case of row-wise partitioned sparse matrix operations, each processes can use the local sparsity pattern to determine which data should be received from other processes, but has insufficient information regarding processes to which data should be sent.  As a result, a crucial first step in every communication phase is determining this pattern of communication.  The creation of this communication pattern requires a \emph{sparse dynamic data exchange} phase (\emph{SDDE}).  The cost of creating the communication pattern through the SDDE often outweighs the cost of subsequent communication unless a number of iterations of communication are performed.

There are two main use cases for sparse dynamic data exchanges: exchanges with constant data sizes, such as determining only the communication pattern and per-message byte count, and variable-sized data exchanges, in which each process must dynamically send a variable amount of data to a subset of other processes.  Cell adaptive mesh refinement applications, such as CELLAR. require an SDDE with constant size messages each time the mesh is altered.  Each process can locally determine the subset of processes to which it must send and what data should be sent to each.  However, an SDDE is needed to determine from which processes each receives, along with the size of each receive.  This SDDE requires constant message sizes as only message sizes must be exchanged.  A large majority of SDDEs, however, require variable-sized exchanges.  Linear solvers typically consist of distributing sparse matrices across processes row-wise, allowing each process to locally determine the receive portion of the communication pattern.  However, in this case, an SDDE is needed for each process to find the subset of processes to which it must send along with which data to send to each.  As a result, indices corresponding to these data values must be sent, requiring each message to be equal to the size of subsequent iterations of communication, and therefore variable.

There are many existing approaches for the sparse dynamic data exchange, each hand implemented within applications that rely on the SDDE.  There are two typical implementations for the sparse dynamic data exchange, which can apply to both variable and constant data sizes.  The personalized method consists of all processes performing an MPI\_Allreduce to determine data exchange sizes before indices are dynamically communicated.  Alternatively, the non-blocking implementation proposed by Torsten et al. dynamically exchanges indices without the need for an initial reduction.  The methods each outperform the other in a subset of sparse dynamic data exchange problems, with performance dependent on communication patterns, data exchange sizes, and process count.  The personalized method outperforms the non-blocking approach in the case of SDDEs with very large message counts, as the allreduce has minimal overhead.  However, SDDEs among a large set of process in which there are few messages are more optimally implemented through the non-blocking algorithm, avoiding the significant overhead of the reduction.

Applications of constant-sided SDDEs, such as Cellar, have further optimized the algorithm with one-sided MPI remote-memory access (RMA) communication.  As all messages are of size \texttt{count}, each process $p$ can use \texttt{MPI\_Put} to directly put its contribution into the destination process's receive buffer at position $p \times \texttt{count}$.  This implementation, however, cannot be naturally extended to the variable-sized SDDE as variable sized data would need to be exchanged after the message sizes are exchanged with RMA.  While the initial RMA step would allow for a standard neighbor collective or point-to-point exchange thereafter, rather than dynamic communication involving probes, very large scales are required for RMA to outperform the personalized and non-blocking approaches.  At these scales, the significant queue search costs, or overhead of matching received data with posted \texttt{MPI\_Irecv}s, cause standard point-to-point exchanges to be more costly than dynamically receiving the first available method with \texttt{MPI\_Probe}.  As a result, the RMA SDDE presented in this paper is only applied to SDDEs with constant-sized messages.


This paper presents optimizations to existing sparse dynamic data exchanges through message aggregation.  Locality-aware optimizations greatly improve the performance and scalability of irregular data exchanges within the context of iterative methods.  Communication within a region, such as intra-socket, greatly outperforms inter-region messages, such as inter-socket or inter-node.  Locality-aware aggregation techniques gather data within a region and minimize the number of messages communicated between regions.  This paper presents two novel locality-aware sparse dynamic data exchange algorithms, which optimize point-to-point communication constraints for large message counts through message aggregation.  Locality-aware variants are presented for both the personalized and the non-blocking SDDE approaches, \emph{achieving up to 20x speedup over existing personalized and non-blocking SDDE methods.}  While the paper does not investigate locality-aware aggregation within the RMA SDDE, the presented aggregation techniques could be naturally extended to one-sided communication.

All SDDE algorithms presented in this paper are implemented within the open-source MPI eXtension library MPI Advance~\cite{MPIAdvance}.  Application programmers can replace current hand-implemented SDDE methods within widely used solvers with the standard API presented within this paper to utilize the presented optimizations.  The authors present scaling studies of the various algorithms as implemented within MPI Advance and analyze the trade-offs between the existing and locality-aware approaches.

The remainder of the paper is outlined as follows. An overview of the sparse dynamic data exchange problem and related works are presented in Section ~\ref{sec:bkgnd}.  A common MPI eXtension (\texttt{MPIX}) API for SDDE algorithms is detailed in Section~\ref{sec:api}.  Existing algorithms for the SDDE are detailed in Section~\ref{sec:impl}, while novel locality-aware extensions are described in Section~\ref{sec:loc}.  Scaling studies are presented in Section~\ref{sec:results}, along with analysis of when each algorithm performs best.  Finally, concluding remarks and future work are discussed in Section~\ref{sec:conc}.

\section{Background}~\label{sec:bkgnd}
The sparse dynamic data exchange problem consists of a set of processes, each of which contains local data and is looking to perform an operation on the global collection of data.  Each process must receive a subset of the global data before it can complete its local portion of the operation.  Initially, every process knows which global data it must receive from other processes.  For instance, consider a sparse matrix operation, in which the sparse matrix and corresponding vectors are partitioned row-wise across all available processes.  Each process must receive data corresponding to each non-zero column within its subset of rows.  However, before data can be exchanged, each process must discover the subset of processes to which it must send all or part of its local data.  The problem is formalized as follows.

\begin{definition}[Sparse Dynamic Data Exchange Problem]
Let $\Sigma:=\{\sigma_1,\dots,\sigma_n\}$ be a collection of processes. 
For each $\sigma_i\in\Sigma$, let there be an associated local data segment $D_i$
and a subset $A_i$ of $\Sigma$ of processes from which $\sigma_i$ needs to receive data from.
The solution to the problem is a communication pattern such that after execution, each $\sigma_i\in\Sigma$ 
holds a copy of the set $\{j\;|\;\sigma_i\in A_j\}$.
\end{definition}

Often, parallel applications rely on the SDDE problem when forming a communication pattern consisting of all processes to which each rank sends and receives data, the sizes of these messages, and indices of the data to be communicated.  There are multiple existing approaches to forming these communication patterns, including the personalized, non-blocking, and RMA approaches, detailed in Section~\ref{sec:impl}.  The performance of existing SDDE approaches consists of overhead costs, such as window synchronization or reduction costs, as well as the cost of point-to-point or one-sided dynamic communication of data.  The trade-offs between existing implementations depend on whether the bottleneck of an SDDE is due to overhead or transfer of data.  Factors that determine this overhead include the number of messages each process communicates, the size of these messages, and the total number of processes.  Furthermore, the performance of the required point-to-point or one-sided communication is dependent on the locality of the messages, MPI implementation, and system architecture.  

\subsection{Parallel Matrix Partitioning}
\begin{figure}[t]
    \centering
    \includegraphics[width=\linewidth]{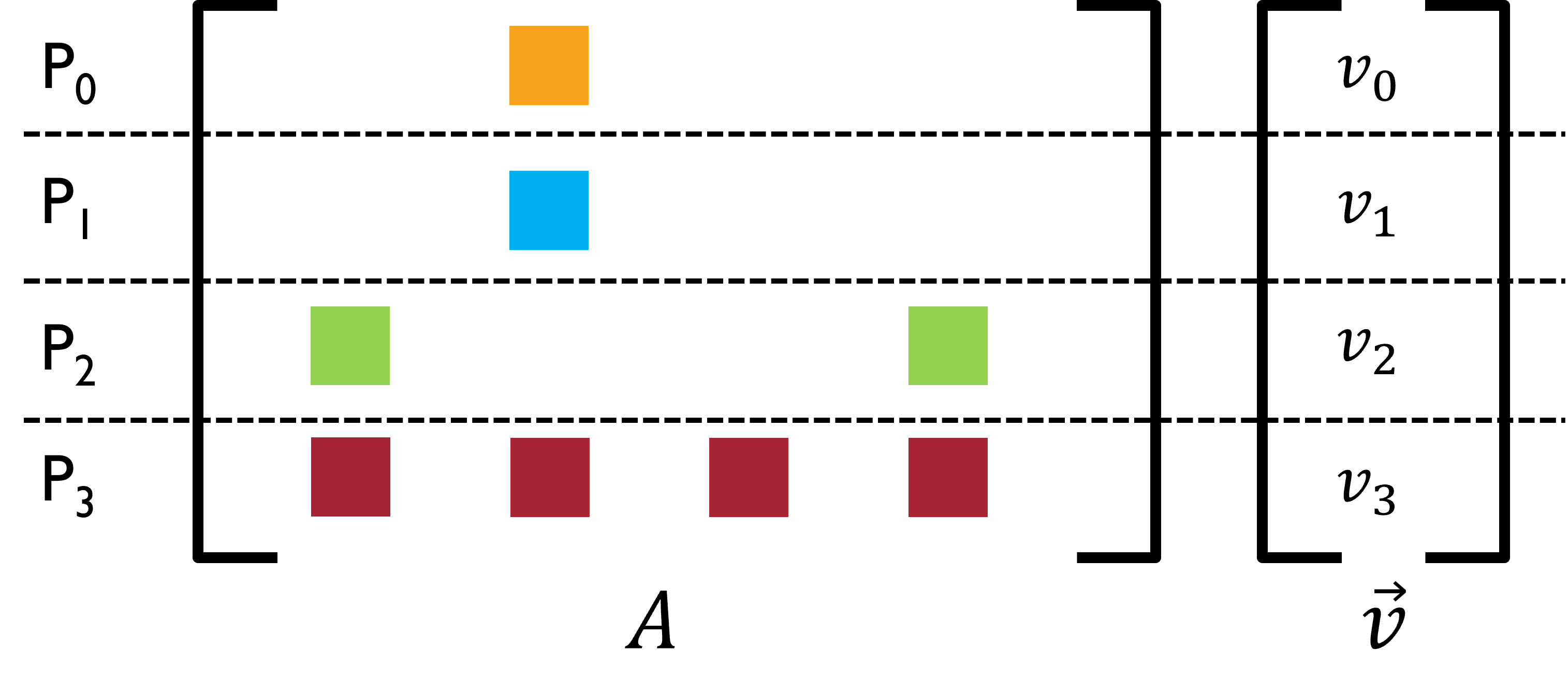}
    \caption{An example of a $4\times4$ matrix and a vector distributed across 4 processes.}
    \label{matrix_partition}
\end{figure}
Sparse matrices are typically partitioned row-wise, so that each process holds a contiguous subset of rows of the matrix along with corresponding vector values, as shown in Figure~\ref{matrix_partition}.  
Consider a system with $n$ rows split row-wise across $p$ processes. Each process holds $\frac{n}{p}$ consecutive rows of the matrix assuming the number of processes divides the number of rows; otherwise, some processes will hold one additional row.  Furthermore each process holds the vector values corresponding to its local columns.  There are other partitioning strategies, such as column-wise or 2.5d partitions~\cite{2dpart_2005,2dpart_2017}.  All SDDE strategies discussed in this paper are applied to row-wise partitions of sparse matrices and extend naturally to column-wise partitioning.  While multi-dimensional partitioning strategies are outside the scope of this paper, the required SDDE problems are similar regardless of the partitioning strategy.


\begin{table*}
    \centering
    \begin{tabular}{| c | c | c || c | c | c |}
        \hline
        \textbf{Send Variable} & \textbf{Input/Output} & \textbf{Definition} & \textbf{Receive Variable} & \textbf{Input/Output} & \textbf{Definition}\\
        \hline
        \hline
        \texttt{send\_nnz} & Input & Number of dynamic sends & \texttt{recv\_nnz} & Input/Output & Number of dynamic sends\\
        \hline
        \texttt{\textcolor{red}{send\_size}} & Input & Size of dynamic sends & \texttt{\textcolor{red}{recv\_size}} & Input/Output & Size of dynamic sends\\
        \hline
        \texttt{dest} & Input & Destinations of messages & \texttt{src} & Output & Destinations of messages\\
        \hline
        \texttt{sendcount\textcolor{red}{s}} & Input & Send per-message count & \texttt{recvcount\textcolor{red}{s}} & Output & Receive per-message count\\
        \hline
        \texttt{\textcolor{red}{sdispls}} & Input & Send displacements & \texttt{\textcolor{red}{rdispls}} & Input & Receive displacements\\
        \hline
        \texttt{sendtype} & Input & Datatype of sends & \texttt{recvtype} & Output & Datatype of receives\\
        \hline
        \texttt{sendvals} & Input & Data to send & \texttt{recvvals} & Output & Data to receive\\
        \hline
    \end{tabular}
    \caption{Variables required in MPIX sparse dynamic data exchange APIs.  Red text indicates variables only required in the variable SDDE, which black variables are required in both APIs.}~\label{table:vars}
\end{table*} 

\subsection{Data Dependencies in Matrix-Vector Multiplications}
Consider the matrix displayed in Figure \ref{matrix_partition}. 
When performing a matrix-vector product $A\vec{v}$ in parallel, each process must retrieve vector entries corresponding to off-process non-zeroes to perform the operation. 
For example, $P_2$ needs vector entries $v_0\text{ and }v_3$ in order to performs its local portion of the matrix-vector product. 
A complete data dependency chart is given by the following figure:
\begin{figure}[h!]
    \centering
    \includegraphics[width=0.7\linewidth]{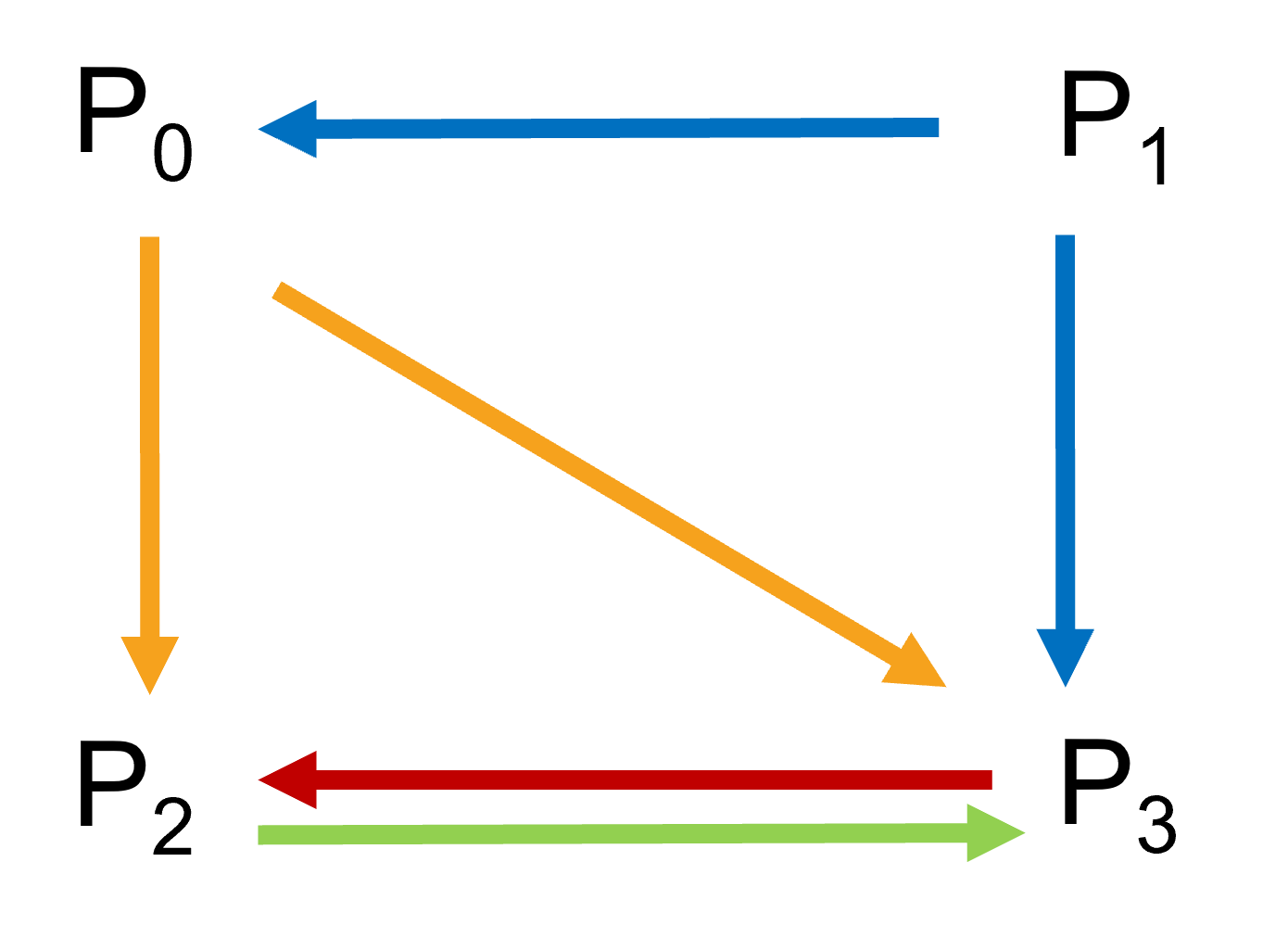}
    \caption{Data dependency chart when performing the matrix vector product illustrated in Figure \ref{matrix_partition}}
    \label{data_dependency}
\end{figure}

As shown in this example, a process does not know in advance which other processes need portions of its data as this communication pattern is dependent on the associated sparsity pattern. 
Therefore, before data can be exchanged, SDDE is required to determine the subset of processes with which each must communicate, along with which data must be communicated.
Communication patterns for irregular data exchanges are typically sparse, with each process communicating with a small subset of the total number of processes, typically $\mathcal{O}(\log(n))$ neighbors.
For more dense communication patterns, all-to-all exchanges are more efficient, removing the needs for the SDDE phase and allowing for optimizations such as pairwise exchange.

\subsection{Related Works}
Irregular communication bottlenecks many parallel applications, including sparse solvers and simulations.  As a result, there has been a significant amount of research into irregular communication optimizations.  However, the majority of research into irregular communication optimizes messages after the communication pattern has been formed.  Iterative point-to-point communication has been optimized within MPI through persistent communication~\cite{persistent2012,persistent2013}, which performs initialization once before reusing the communicator, amortizing setup costs over all iterations.  MPI 4.0 provides partition communication~\cite{partitioned2019,partitioned2021}, an optimization for multithreaded applications that allows each thread to send a portion of a message when ready.

Many architecture-aware optimizations for irregular and collective communication utilize locality to reduce communication costs.  Hierarchical communication reduces inter-node communication costs by utilizing only one or a small number of MPI processes per node during inter-node communication~\cite{hierarchical1,hierarchical2}.  Locality-aware aggregation minimizes the number and size of non-local messages, such as inter-node, through local, such as intra-node, aggregation of data~\cite{nodeawarespmv,nodeawareamg, HidayetogluLocAwareGPU}.  Similar node-aware aggregation strategies, such as YGM, have been applied to graph operations~\cite{ygm}.  Multilane algorithms minimize the size of inter-node messages, having each process per node send an equal and minimal amount of data~\cite{multilane}.


Communication constraints can be minimized through optimal partitions.  Rather than a standard row-wise partition, graph or hypergraph partitioners minimize edge cuts between partitions~\cite{partitioning}.  Each partition is then assigned to a single process or node, reducing the amount of data needed to be exchanged throughout applications.  While graph partitioners can be used in combination with sparse dynamic data exchanges, the use of these methods is outside the scope of this paper due to their large overheads.




\section{MPI Extension APIs}
A large variety of sparse dynamic data exchange implementations  are used throughout existing parallel applications, with each application required to determine the optimal method for their program.  This section presents the API that the authors have added to an open-source MPI Extension library, MPI Advance.  Existing methods and novel optimizations are added within the provided APIs, allowing an existing application to access any of the available optimizations rather than hand optimize.  Further, the API allows for future research in which the optimal algorithm can be selected dynamically.  Note that sparse data exchanges exist within MPI through neighbor collectives, but these methods do not allow for sparse \emph{dynamic} data exchanges.  

Sparse dynamic data exchange algorithms can be represented with two separate APIs: \texttt{MPIX\_Alltoall\_crs} and \texttt{MPIX\_Alltoallv\_crs}.  All variables required for the APIs are defined in Table~\ref{table:vars}.

The \texttt{MPIX\_Alltoall\_crs} method allows for dynamically exchanging constant data sizes with a sparse set of neighbors.  One use case of this method occurs when processes know all data that must be sent to other processes, but have no knowledge of the receive portion of the communication pattern.  This type of SDDE algorithm occurs within cell-based adaptive mesh refinement solvers, such as CELLAR~\cite{xrage_communication}.

The \texttt{MPIX\_Alltoall\_crs} API is defined in Figure~\ref{fig:alltoall_api}.  
\begin{figure}[ht!]
\begin{lstlisting}
int MPIX_Alltoall_crs(int send_nnz, 
        int* dest, int sendcount, 
        MPI_Datatype sendtype, 
        void* sendvals, int* recv_nnz, 
        int* src, int recvcount, 
        MPI_Datatype recvtype, 
        void* recvvals, MPIX_Info* xinfo, 
        MPIX_Comm* xcomm)
\end{lstlisting}
\caption{\texttt{MPIX\_Alltoall\_crs} API}~\label{fig:alltoall_api}
\end{figure}
This method takes the send portion of the SDDE as input and returns received data.  The variable \texttt{recv\_nnz} is both input and output, allowing the user to input this information if it is already known rather than requiring the method to redetermine number of receives.  Note the API is following current MPI standard requirements, and as a result \texttt{recvvals} must be allocated, potentially to some upper-bound, before the method is instantiated.

The \texttt{MPIX\_Alltoallv\_crs} method provides an interface for dynamically exchanging variable sized data with a sparse set of neighbors.  This method is required for any distributed sparse matrix operation, in which processes know all data that must be received, but have no knowledge of processes to which they must send or what data must be send to each. As a result, a variable sized message must be exchanged containing this boundary information.  Typically, this variable SDDE is used to initially form a communication package to then be used during each future sparse matrix operation, during which communication is typically implemented either through point-to-point messages or neighborhood collectives.  Variable SDDEs exist within many widely used solvers, including Hypre BoomerAMG.

The \texttt{MPIX\_Alltoallv\_crs} API is defined in Figure~\ref{fig:alltoallv_api}. 
\begin{figure}[ht!]
\begin{lstlisting}
int MPIX_Alltoallv_crs(int send_nnz, 
        int send_size, int* dest, 
        int* sendcounts, int* sdispls, 
        MPI_Datatype sendtype, 
        void* sendvals, int* recv_nnz, 
        int* recv_size, int* src, 
        int* recvcounts, int* rdispls, 
        MPI_Datatype recvtype, 
        void* recvvals, MPIX_Info* xinfo, 
        MPIX_Comm* comm)
\end{lstlisting}
\caption{\texttt{MPIX\_Alltoallv\_crs} API}~\label{fig:alltoallv_api}
\end{figure}
Similar to the non-variable API, this method takes the send portion of the variable SDDE as input and returns received data.  The parameters \texttt{recv\_nnz} and \texttt{recv\_size} are both input and output, allowing the user to input both receive count and/or receive size if either are known.  Due to MPI standard requirements, this method also requires all pointers to be returned by the method to be allocated before they are passed to the method.  This includes \texttt{recvcounts} and \texttt{rdispls}, which each need to be allocated to hold \texttt{recv\_nnz} integers at mininum, as well as \texttt{recvvals}, which needs to be allocated to hold at least \texttt{recv\_size} variables of size \texttt{recvtype}.~\label{sec:api}

\section{Implementations}~\label{sec:impl}
There are a number of existing methods for determining the communication pattern of a given irregular data exchange.  The personalized and non-blocking methods, described in detail below, are widely used throughout existing parallel applications.  Furthermore, an RMA-based optimization, as implemented within Cellar, allows for \texttt{MPIX\_Alltoall\_crs} communication to be performed with one-sided communication.  \textbf{All algorithms presented in this section, except for the RMA method, can be applied to both the \texttt{MPIX\_Alltoall\_crs} and \texttt{MPIX\_Alltoallv\_crs} methods, with minor changes required for the variable-sized SDDE, as highlighted in red.}

\subsection{Personalized Method}
The standard personalized method, utilized in a number of irregular codebases and described in~\cite{torsten1}, is detailed in Algorithm~\ref{alg:standard}.  
\begin{algorithm2e}[ht!]
  \DontPrintSemicolon%
  \KwIn{\texttt{rank} \tcc*{Process ID}
        \texttt{args} \tcc*{\texttt{MPIX\_Alltoall\textcolor{red}{v}\_crs} Arguments}
        }

  \BlankLine%
\BlankLine%

  $\texttt{ctr}\gets0$\;
  \For{$i\gets0$ \KwTo \texttt{send\_nnz}}{
      \texttt{proc} = \texttt{dest}[$i$]\;
      
      \texttt{size} = \texttt{sendcount\textcolor{red}{s}}\textcolor{red}{[$i$]}\;

      \texttt{sizes}[\texttt{proc}] = \texttt{size}\;
      
      \texttt{MPI\_Isend} \texttt{size} at  \texttt{sendvals}[\texttt{ctr}] to \texttt{proc}\;

      $\texttt{ctr} \gets \texttt{ctr} + \texttt{size}$\;
  }
  \BlankLine%
  MPI\_Allreduce(\texttt{sizes})
\BlankLine%
  \For{$i\gets0$ \KwTo \texttt{sizes}[\texttt{rank}]}{
      Probe for message and receive dynamically\;
  }
  
  Wait for all sends to complete\;
  
    \caption{Personalized}\label{alg:standard}
\end{algorithm2e}
The personalized protocol completes a sparse dynamic data exchange by performing an MPI\_Allreduce to gather either the total number of size of messages that every process will receive during the dynamic communication phase.  Data to be exchanged is then communicated with non-blocking sends.  Finally, each process dynamically receives all messages sent to it using MPI\_Probe.

When process counts are large, the overhead of the \texttt{MPI\_Allreduce} increases.  However, as message count increases, the reduction overhead is quickly outweighed by the dynamic exchange of data.  The use of probes requires received data to be added to the unexpected message queue, further reducing performance for large messages.  However, the MPI\_Allreduce can yield large benefits as it allows all communication structures to be allocated before the dynamic communication step.

\subsection{Non-blocking Method}
The non-blocking method, presented in Torsten et al.~\cite{torsten1} and detailed in Algorithm~\ref{alg:torsten}, is a modification of the standard algorithm which avoids the overhead and collective synchronization required by the \texttt{MPI\_Allreduce}.

\begin{algorithm2e}[ht!]
  \DontPrintSemicolon%
  \KwIn{\texttt{rank} \tcc*{Process ID}
        \texttt{args} \tcc*{\texttt{MPIX\_Alltoall\textcolor{red}{v}\_crs} Arguments}
    }

  \BlankLine%
\BlankLine%

  $\texttt{ctr}\gets0$\;
  \For{$i\gets0$ \KwTo \texttt{send\_nnz}}{
      \texttt{proc} = \texttt{dest}[$i$]\;
      
      \texttt{size} = \texttt{sendcount\textcolor{red}{s}}\textcolor{red}{[$i$]}\;
      
      \texttt{MPI\_Isend} \texttt{size} at  \texttt{sendvals}[\texttt{ctr}] to \texttt{proc}\;

      $\texttt{ctr} \gets \texttt{ctr} + \texttt{size}$\;
  }

\BlankLine%
  \While{All sends have not completed}{
      \If {Non-blocking probe finds a message}{
          Dynamically receive message\;
      }
  }

  Non-blocking barrier\;

  \While{Non-blocking barrier has not completed}{
      \If {Non-blocking probe finds a message}{
          Dynamically receive message\;
      }
  }
  
    \caption{Non-blocking}\label{alg:torsten}
\end{algorithm2e}

In this algorithm, each process $p$ first sends non-blocking synchronous sends to every process to which it must communicate.  A synchronous send is only considered complete when the destination process posts the associated receive.
Each process dynamically receives messages until the synchronous sends have been completed on all processes.  This is accomplished through MPI\_Iprobe, which checks if a message is available, and only then is data received.  While checking for messages, each process also tests its synchronous sends for completion 
Once all message sent from a process have been received, that process calls a non-blocking barrier to indicate that it is finished sending.  This process continues probing for messages until all processes have reached the barrier, at which point all processes' sends are received and dynamic communication is completed.

While the non-blocking algorithm improves over the personalized method for large process counts by reducing synchronization and cost associated with the MPI\_Allreduce, communication pattern structures must now be dynamically allocated.  Furthermore, dynamic communication and unexpected messages remain a dominant cost of the method.

\subsection{RMA Constant-Size SDDE}
One-sided optimizations exist for sparse dynamic data exchanges in which all data is of a constant size, as implemented in Cellar and detailed in Algorithm~\ref{alg:rma}.
\begin{algorithm2e}[ht!]
  \DontPrintSemicolon%
  \KwIn{\texttt{rank} \tcc*{Process ID}
        \texttt{args} \tcc*{\texttt{MPIX\_Alltoall\_crs} Arguments}
        }

  \BlankLine%
\BlankLine%

  \textnormal{Create window}\;

  \textnormal{Synchronize on window}\;

  $\texttt{proc\_vals}[\texttt{num\_procs} \times \texttt{sendcount}]$\;

  $\texttt{ctr}\gets0$\;
  \For{$i\gets0$ \KwTo \texttt{send\_nnz}}{
      $\texttt{proc} \gets \texttt{dest}[i]$\;
      
      $\texttt{orig\_address} \gets \texttt{sendvals}[\texttt{ctr}]$\;
      
      $\texttt{dest\_address} \gets \texttt{proc\_vals}[\texttt{rank}] \times \texttt{sendcount}$\;
      
      \texttt{MPI\_Put} \texttt{sendcount} values from \texttt{orig\_address} to \texttt{dest\_address}\;
      
      $\texttt{ctr} \gets \texttt{ctr} + \texttt{sendcount}$\;
  }

    \textnormal{Synchronize on window}\;

     \For{$i\gets0$ \KwTo \texttt{num\_procs}}{
    $\texttt{}{val} \gets \texttt{proc\_vals}[i \times \texttt{sendcount}]$\;
        \If{\texttt{val}}{
            $\texttt{recvvals}[\texttt{recv\_nnz}++] \gets \texttt{val}$\;
        }
     }

    \caption{RMA (\texttt{MPIX\_Alltoall\_crs} only)}\label{alg:rma}
\end{algorithm2e}
Each process firsts allocates a shared memory window with enough space for all processes to add \texttt{sendcount} values.  Then, each process can independently use \texttt{MPI\_Put} to move the required data to position $\texttt{rank} \cdot \texttt{sendcount}$ on each corresponding process.  Finally, each process can move received data from the window into the \texttt{recvvals} array.

The one-sided approach requires window creation, but this can be amortized over the cost of the application, and can potentially be reused within any subsequent one-sided communication.  There is also a synchronization required to use a shared memory window.  The author's currently use \texttt{MPI\_Fence} for this, but more optimized forms of synchronization, such as locks, could be explored.  RMA sparse dynamic data exchanges allow for all data to be exchanged without any dynamic MPI communication.  However, the method does not naturally extend to variable-sized dynamic data exchanges.  It can be used to determine communication sizes, followed a standard data exchange to avoid dynamic messages, but the authors were unable to find a case where this would outperform other methods due to large matching costs.  It is possible, however, locality-aware data exchanges could improve the performance of the subsequent data exchange, but this optimization is outside the scope of this paper.

\subsection{Locality-Aware Method}~\label{sec:loc}
At large scales, the cost of both the personalized and non-blocking methods are bottlenecked by the point-to-point communication requirements.  Irregular communication has been extensively modeled and analyzed, showing large costs associated with high inter-socket message counts due to lower bandwidth than intra-socket messages, injection bandwidth limits, queue search (or matching) costs, and network contention.  This irregular communication can be optimized through locality-awareness, aggregating messages within a region, such as a socket, to minimize the number of messages communicated between regions.  This greatly reduces the number of times inter-region latency is incurred along with associated queue search costs.  While other locality-aware techniques remove duplicate values to reduce inter-region message sizes, this paper only concatenates messages as the overhead of determining duplicate values would outweigh benefits of reduced message size within a single iteration of communication, such as required during the SDDE problem.

\begin{figure*}
     \centering
     \hfill
     \begin{subfigure}[ht!]{0.48\textwidth}
         \centering
         \includegraphics[width=\linewidth]{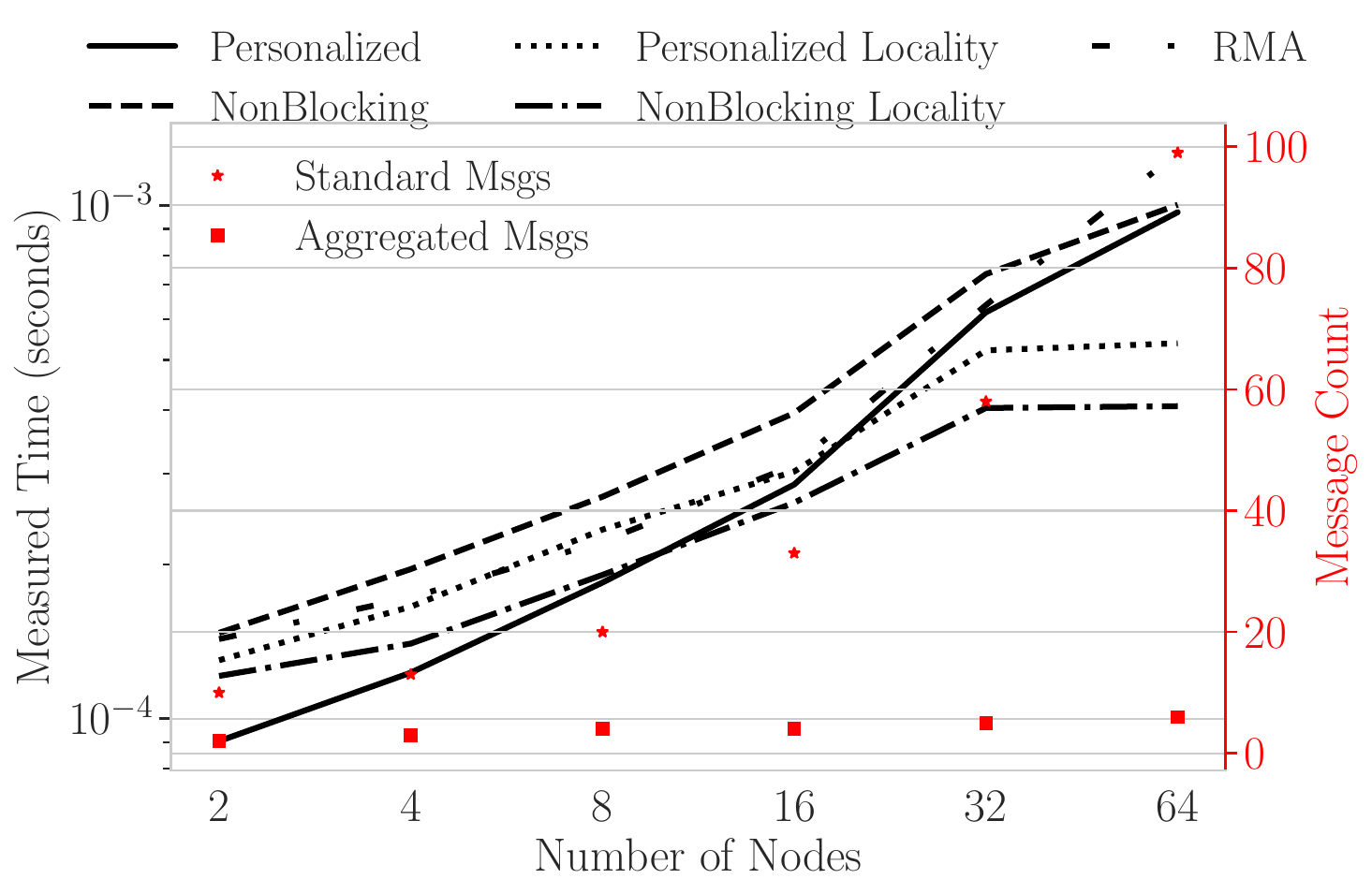}
         \caption{dielFilterV2clx}
     \end{subfigure}
     \hfill
     \begin{subfigure}[ht!]{0.48\textwidth}
         \centering
         \includegraphics[width=\linewidth]{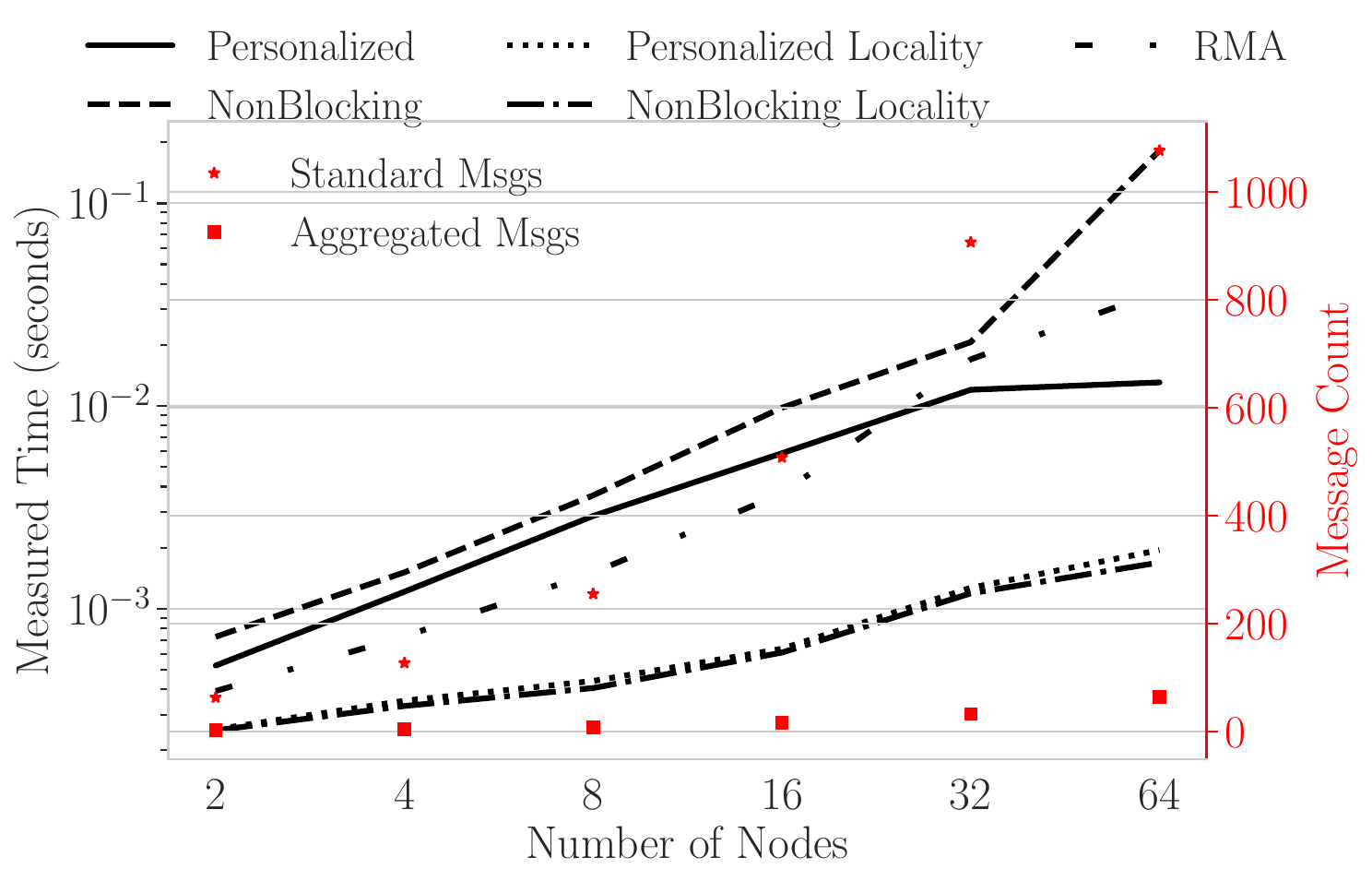}
         \caption{germany\_osm}
     \end{subfigure}
     \hfill
     \begin{subfigure}[ht!]{0.48\textwidth}
         \centering
         \includegraphics[width=\linewidth]{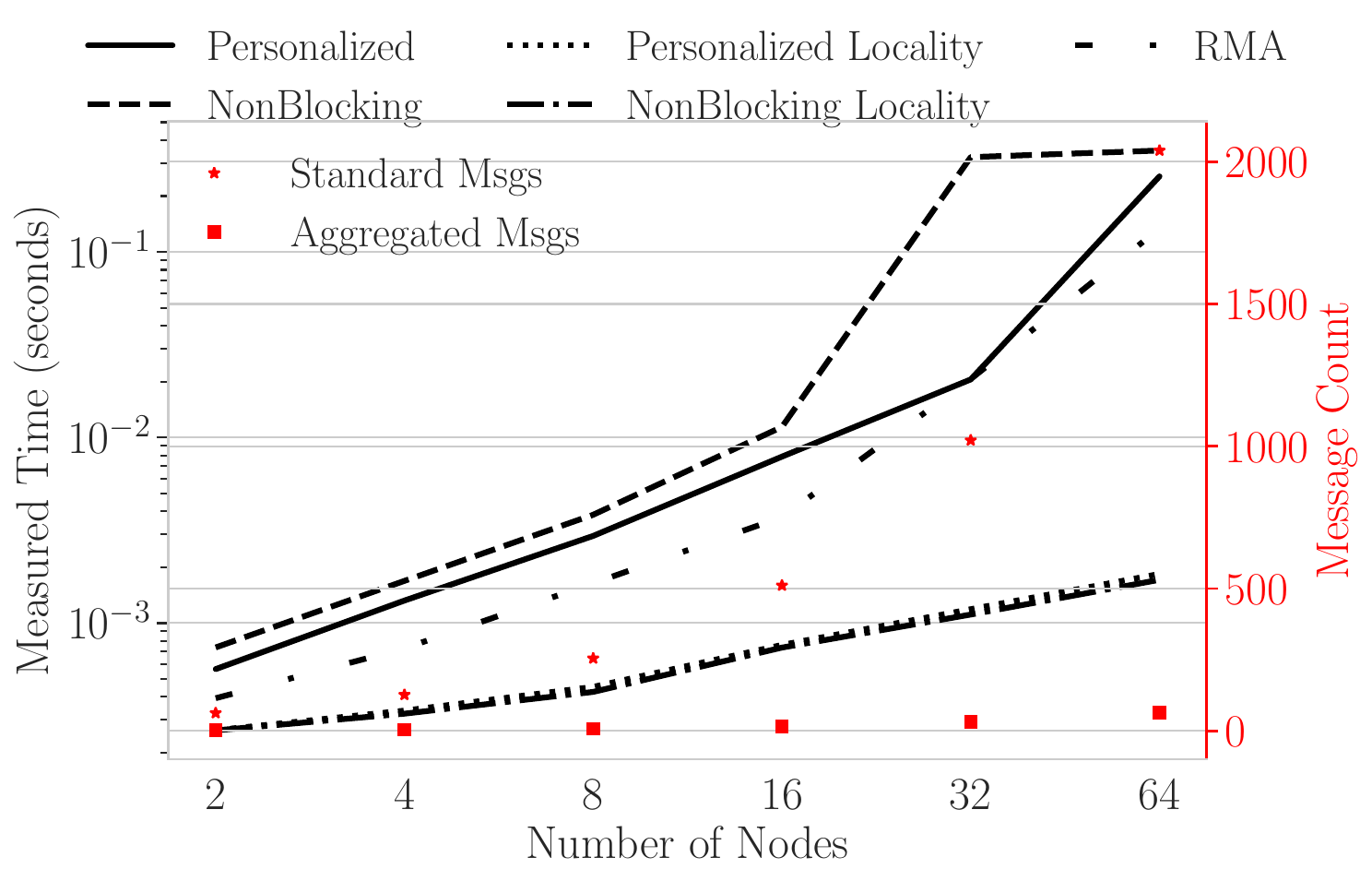}
         \caption{human\_gene1}
     \end{subfigure}
     \hfill
     \begin{subfigure}[ht!]{0.48\textwidth}
         \centering
         \includegraphics[width=\linewidth]{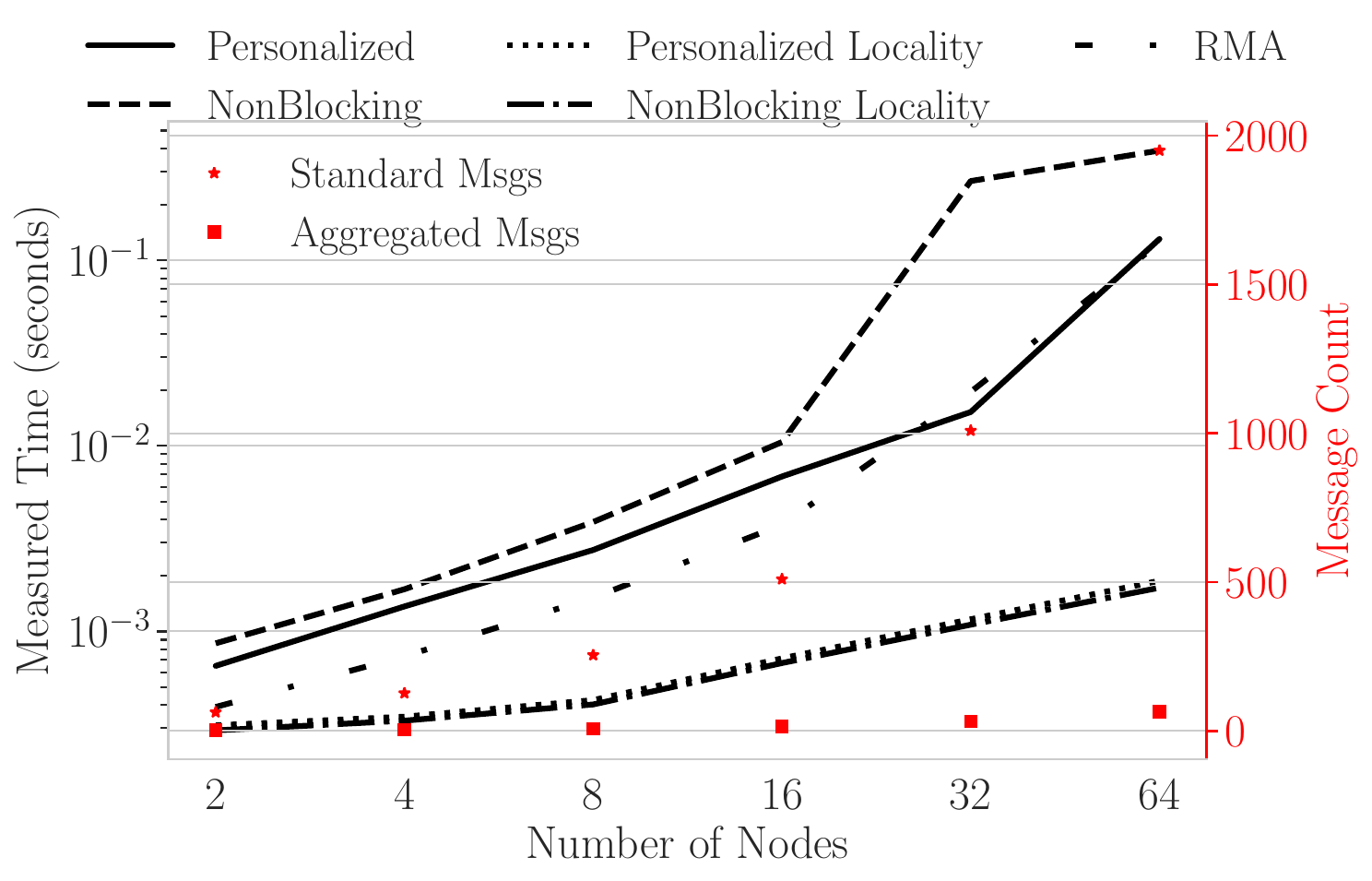}
         \caption{NLR}
     \end{subfigure}
        \caption{The cost of the various \texttt{MPIX\_\textbf{alltoall}\_crs} methods using \textbf{Mvapich2}.}
        \label{fig:alltoall_mvapich}
\end{figure*}
\begin{figure*}
     \centering
     \begin{subfigure}[ht!]{0.48\textwidth}
         \centering
         \includegraphics[width=\linewidth]{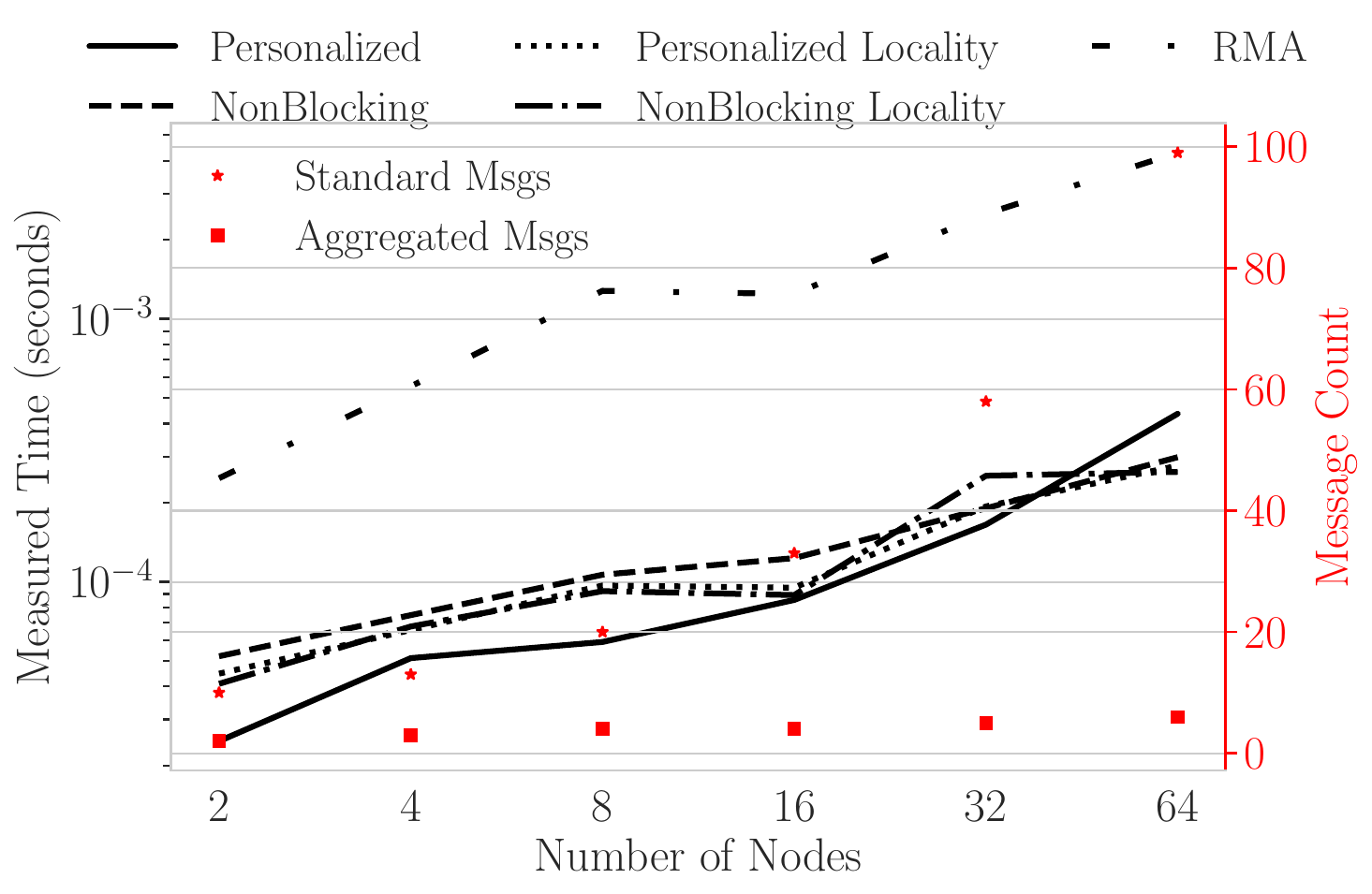}
         \caption{dielFilterV2clx}
     \end{subfigure}
     \hfill
     \begin{subfigure}[ht!]{0.48\textwidth}
         \centering
         \includegraphics[width=\linewidth]{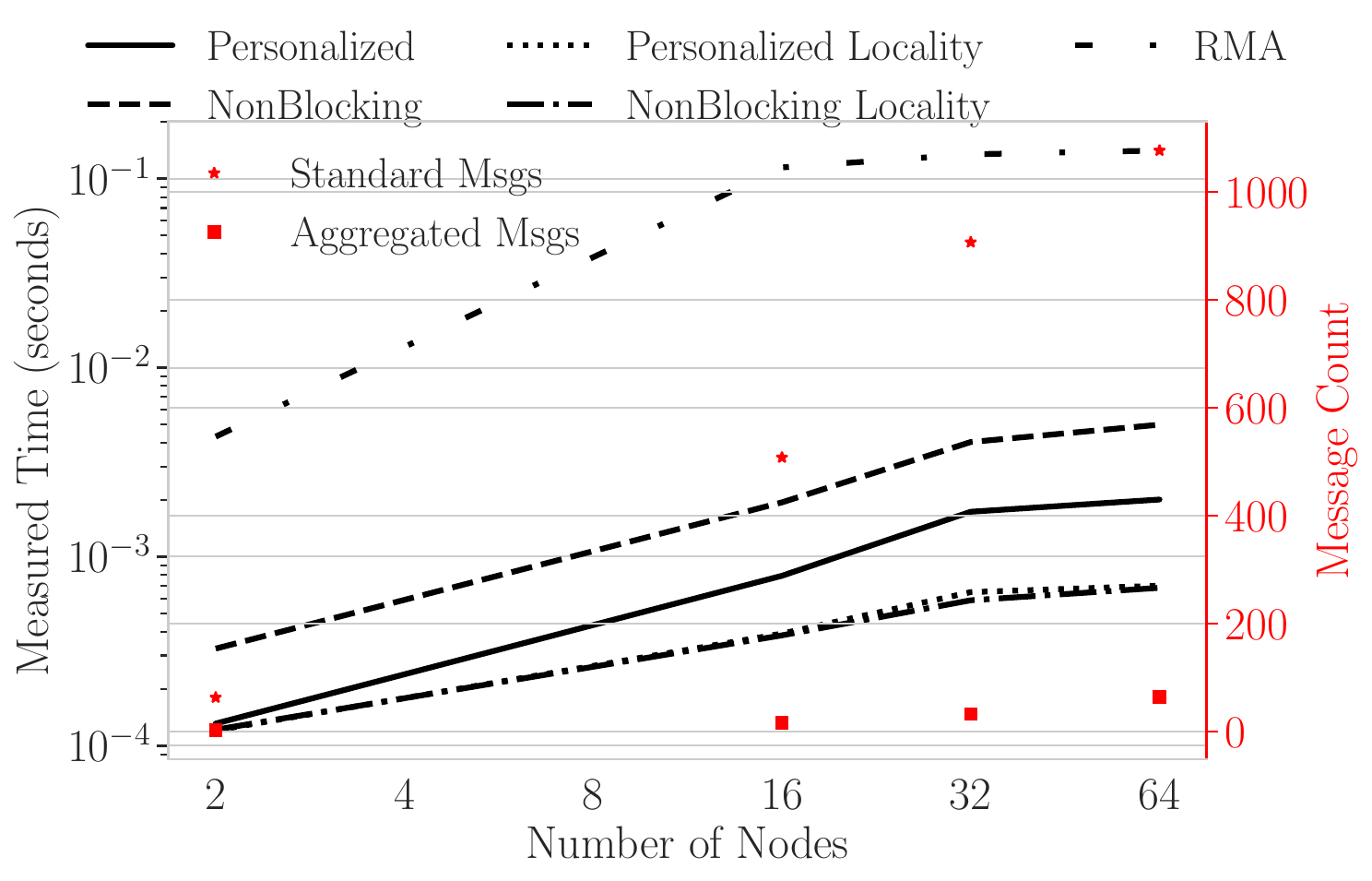}
         \caption{germany\_osm}
     \end{subfigure}
     \hfill
     \begin{subfigure}[ht!]{0.48\textwidth}
         \centering
         \includegraphics[width=\linewidth]{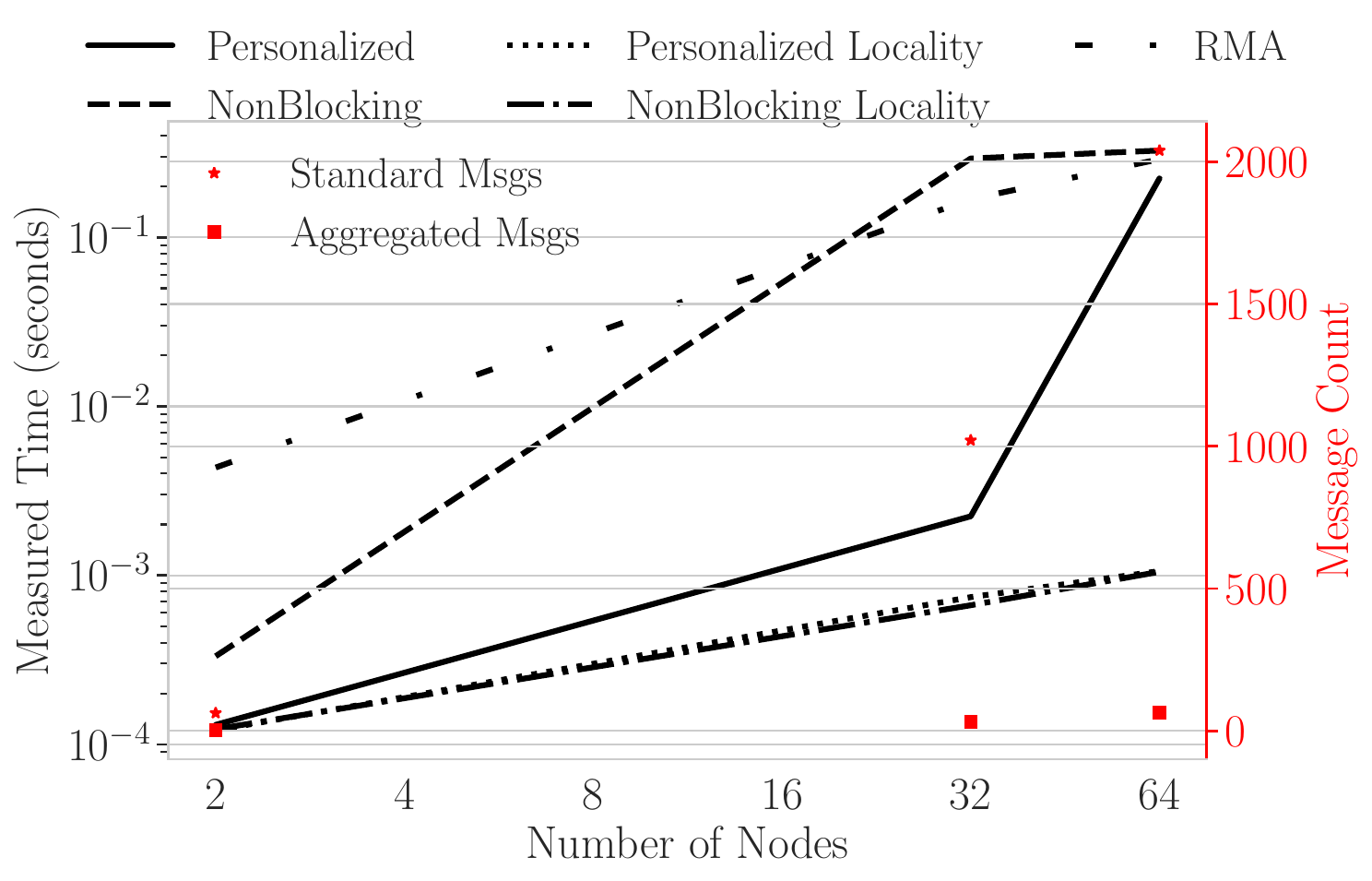}
         \caption{human\_gene1}
     \end{subfigure}
     \hfill
     \begin{subfigure}[ht!]{0.48\textwidth}
         \centering
         \includegraphics[width=\linewidth]{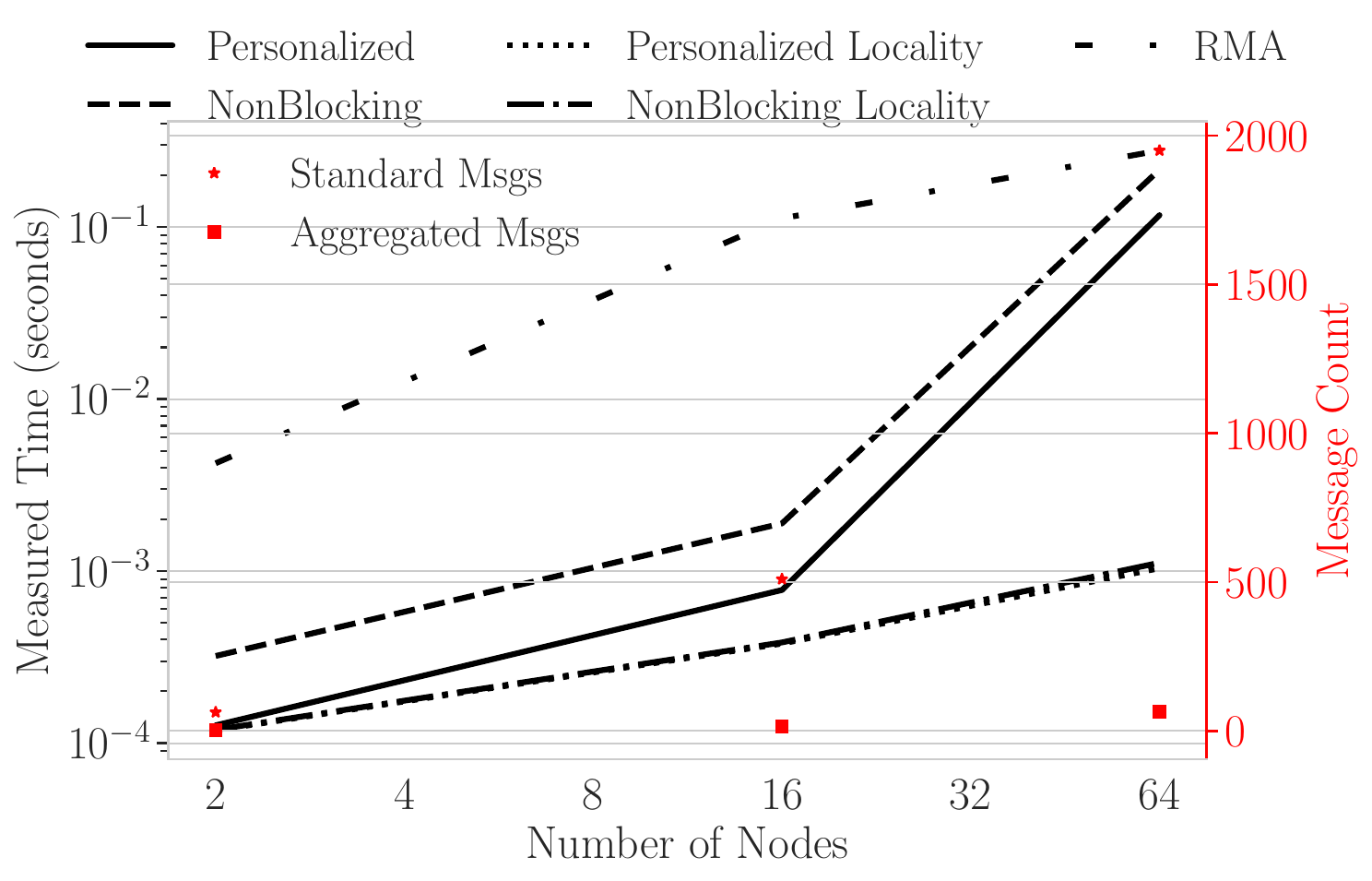}
         \caption{NLR}
     \end{subfigure}
        \caption{The cost of the various \texttt{MPIX\_\textbf{alltoall}\_crs} methods using \textbf{OpenMPI}.}
        \label{fig:alltoall_openmpi}
\end{figure*}
\begin{figure*}
     \centering
     \begin{subfigure}[ht!]{0.48\textwidth}
         \centering
         \includegraphics[width=\linewidth]{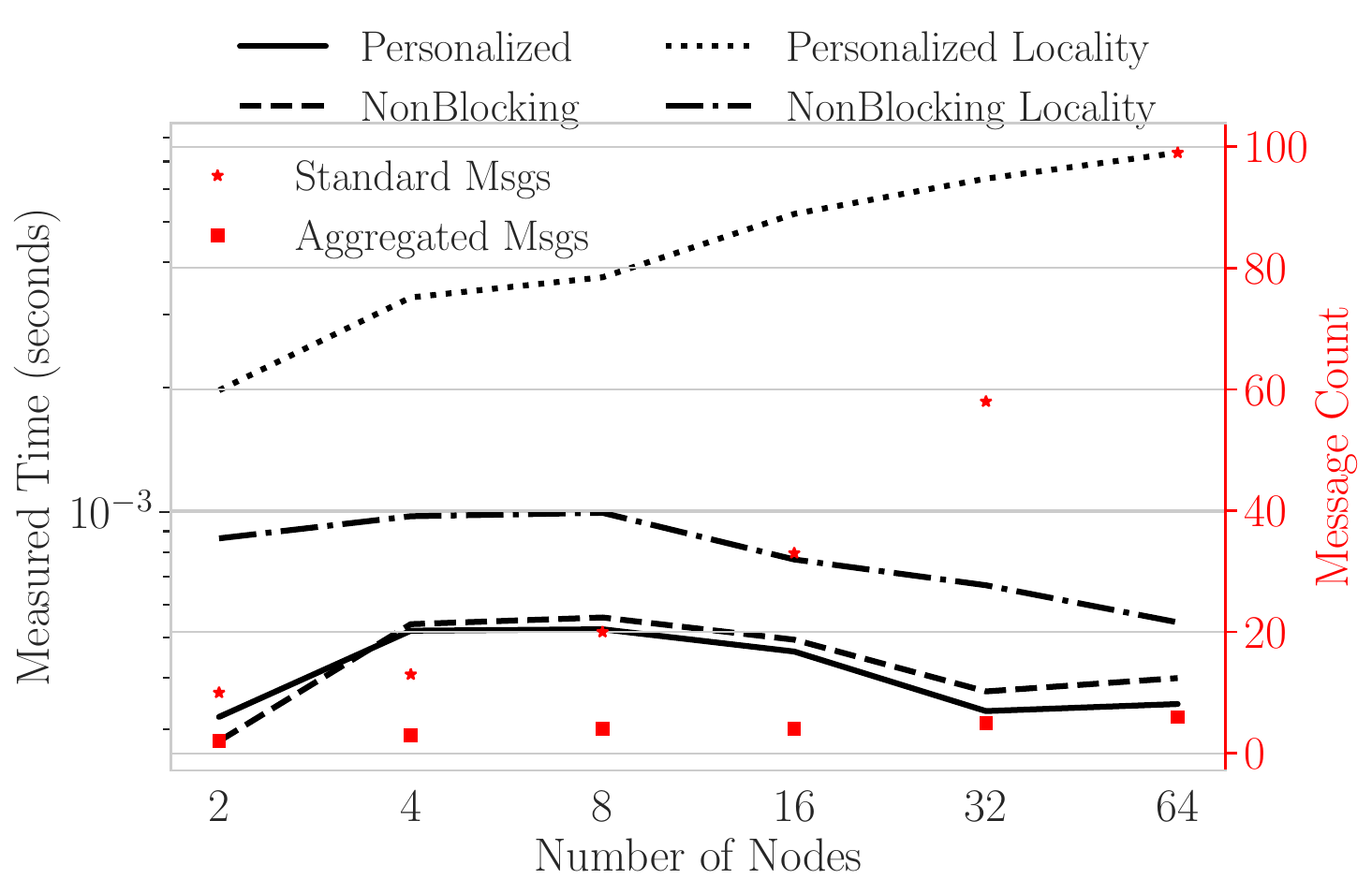}
         \caption{dielFilterV2clx}
     \end{subfigure}
     \hfill
     \begin{subfigure}[ht!]{0.48\textwidth}
         \centering
         \includegraphics[width=\linewidth]{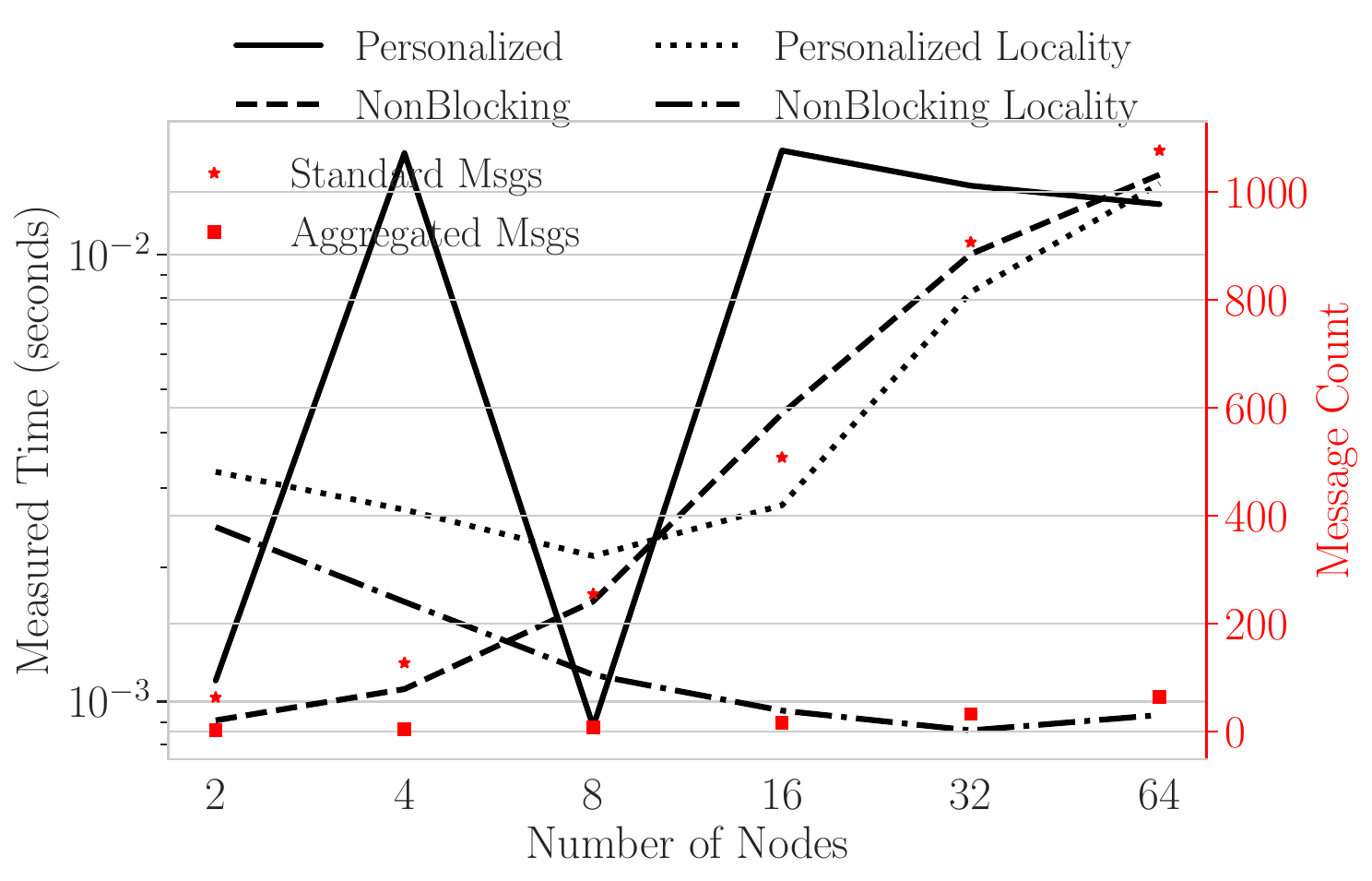}
         \caption{germany\_osm}
     \end{subfigure}
     \hfill
     \begin{subfigure}[ht!]{0.48\textwidth}
         \centering
         \includegraphics[width=\linewidth]{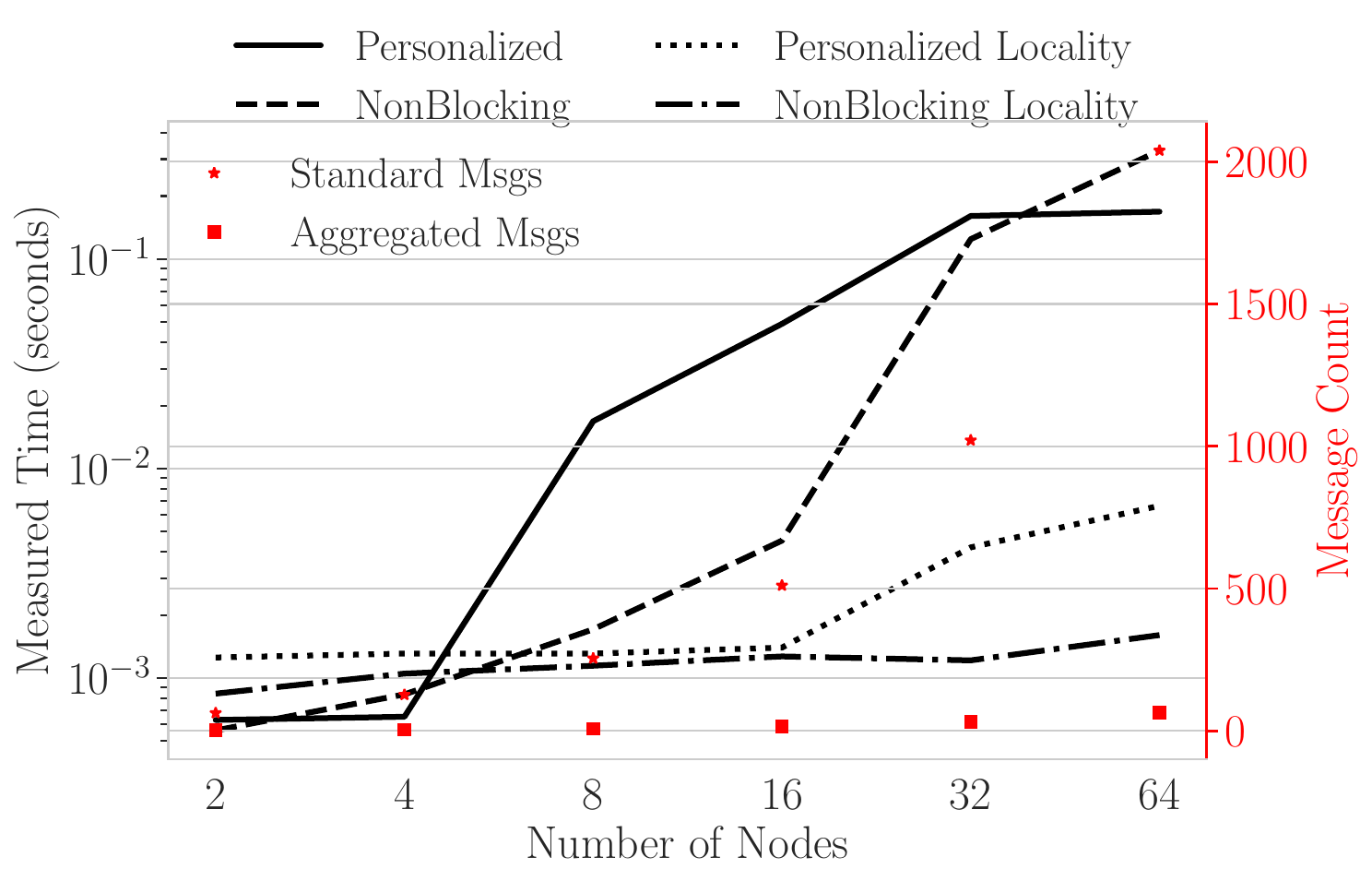}
         \caption{human\_gene1}
     \end{subfigure}
     \hfill
     \begin{subfigure}[ht!]{0.48\textwidth}
         \centering
         \includegraphics[width=\linewidth]{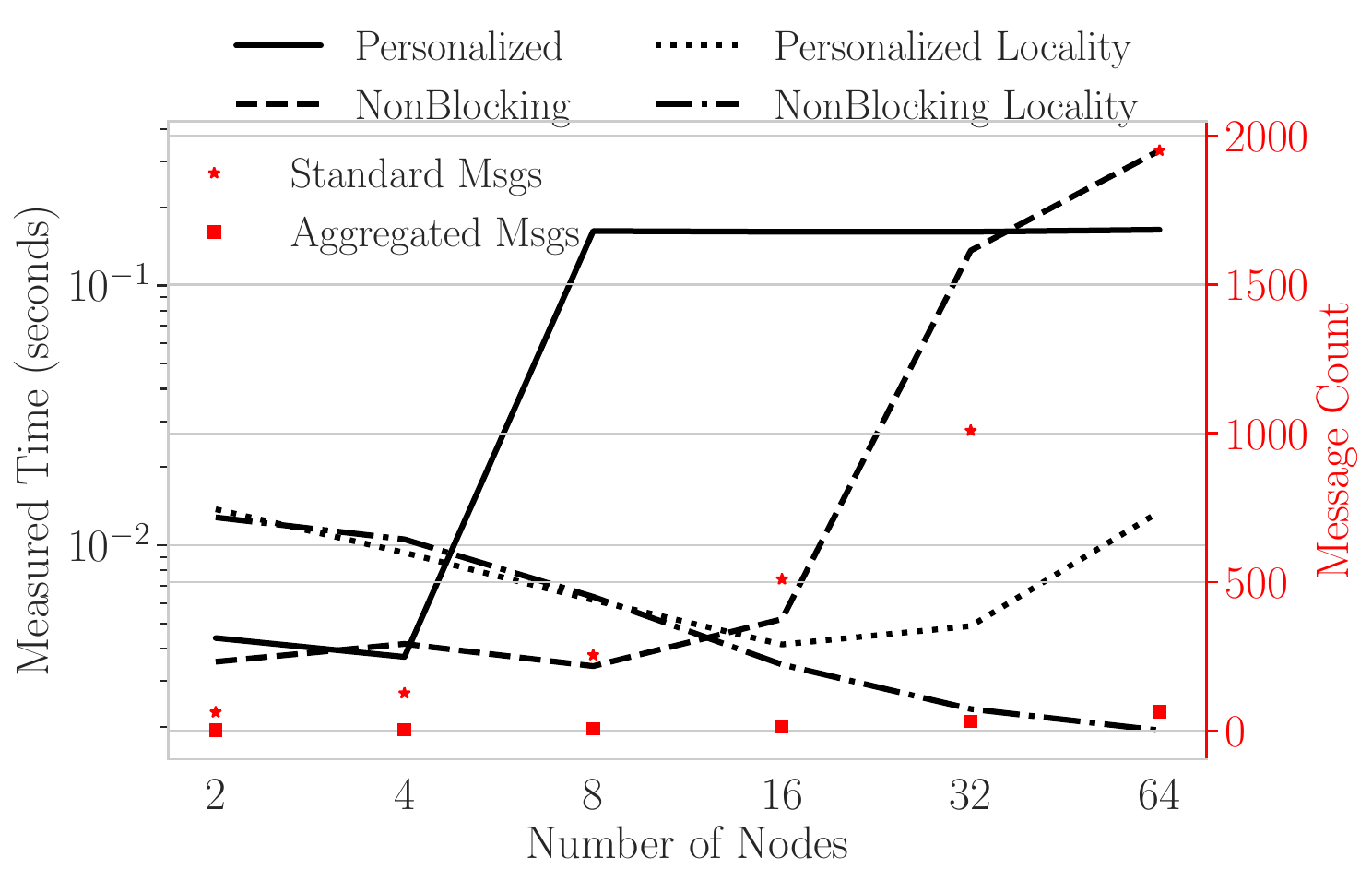}
         \caption{NLR}
     \end{subfigure}
        \caption{The cost of the various \texttt{MPIX\_\textbf{alltoallv}\_crs} methods using \textbf{Mvapich2}.}
        \label{fig:alltoallv_mvapich}
\end{figure*}
\begin{figure*}
     \centering
     \begin{subfigure}[ht!]{0.48\textwidth}
         \centering
         \includegraphics[width=\linewidth]{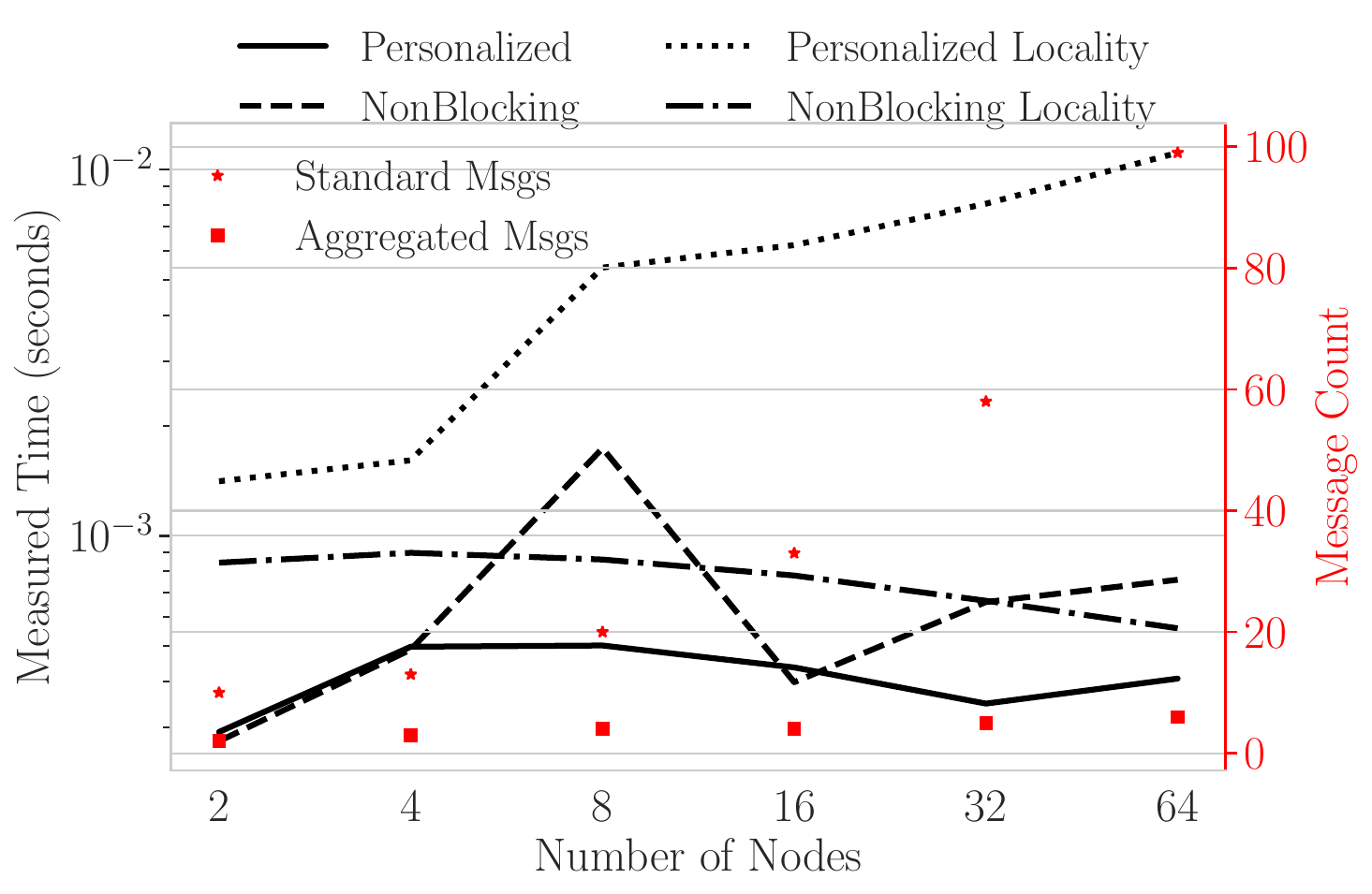}
         \caption{dielFilterV2clx}
     \end{subfigure}
     \hfill
     \begin{subfigure}[ht!]{0.48\textwidth}
         \centering
         \includegraphics[width=\linewidth]{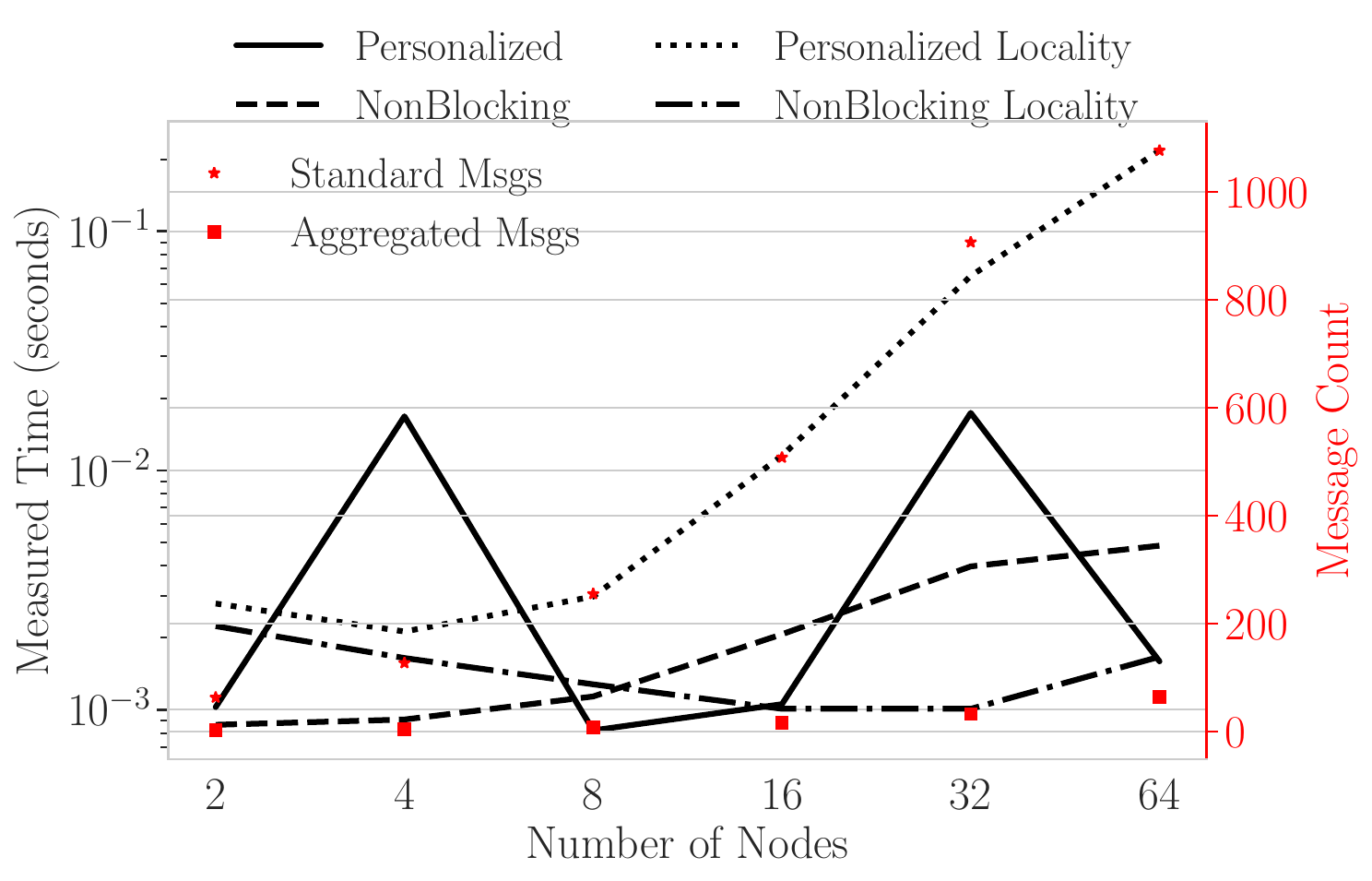}
         \caption{germany\_osm}
     \end{subfigure}
     \hfill
     \begin{subfigure}[ht!]{0.48\textwidth}
         \centering
         \includegraphics[width=\linewidth]{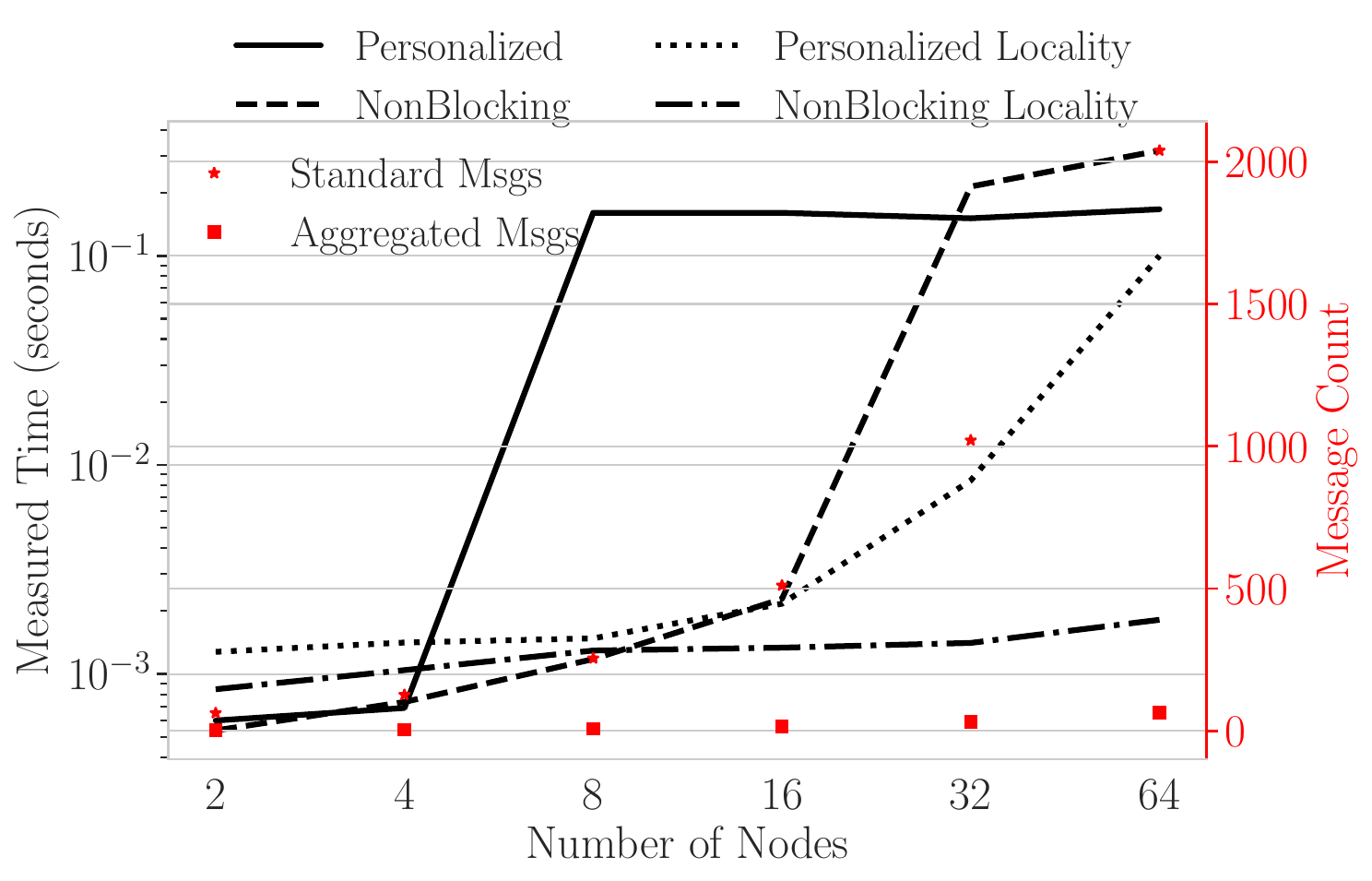}
         \caption{human\_gene1}
     \end{subfigure}
     \begin{subfigure}[ht!]{0.48\textwidth}
         \centering
         \includegraphics[width=\linewidth]{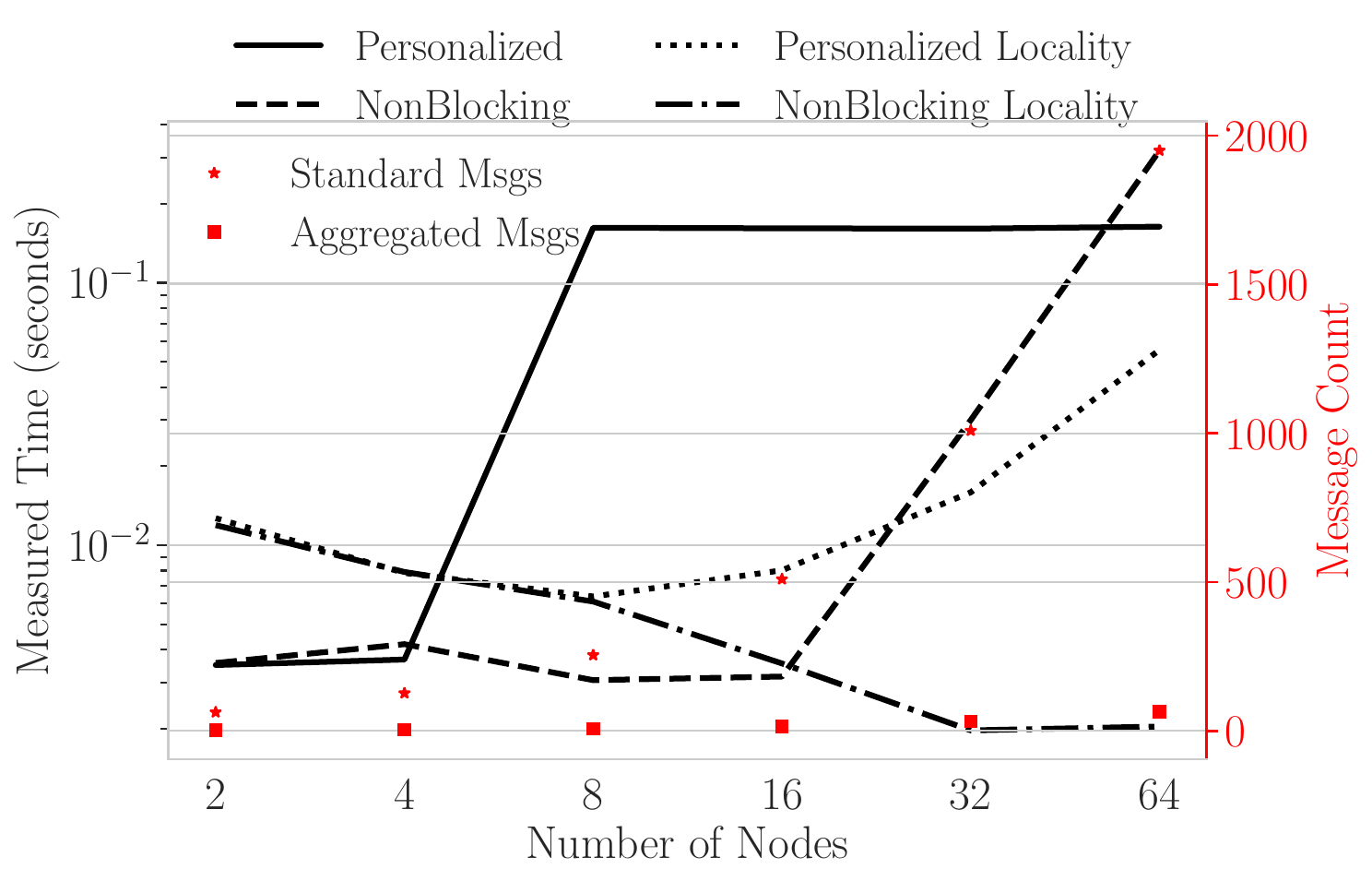}
         \caption{NLR}
     \end{subfigure}
        \caption{The cost of the various \texttt{MPIX\_\textbf{alltoallv}\_crs} methods using \textbf{OpenMPI}.}
        \label{fig:alltoallv_openmpi}
\end{figure*}

This paper presents locality-aware aggregation for both the personalized and non-blocking sparse dynamic data exchanges, as shown in Algorithms~\ref{alg:personalized_loc} and~\ref{alg:n onblocking_loc}, respectively.
\begin{algorithm2e}[ht!]
  \DontPrintSemicolon%
  \KwIn{\texttt{rank} \tcc*{Process ID}
        \texttt{args} \tcc*{\texttt{MPIX\_Alltoall\textcolor{red}{v}\_crs} Arguments}
        \texttt{local\_rank} \tcc*{Local rank of process within region}
        \texttt{region\_size} \tcc*{Number of processes per region}
        }

  \BlankLine%
  \BlankLine%

  $\texttt{ctr}\gets0$\;
  \For{$i\gets0$ \KwTo \texttt{send\_nnz}}{
      \texttt{proc} = \texttt{dest}[$i$]\;
      \texttt{region} = \texttt{get\_region(proc)}\;
      \texttt{size} = \texttt{sendcount\textcolor{red}{s}}\textcolor{red}{[$i$]}\;
      
      Copy \texttt{proc}, \textcolor{red}{size}, and \texttt{sendvals}[\texttt{ctr}:\texttt{ctr}+\texttt{size}] into \texttt{buf}[\texttt{region}]\;

      $\texttt{ctr} \gets \texttt{ctr} + \texttt{size}$\;
  }

  \ForEach{\texttt{region} to which \texttt{rank} sends}{
     $\texttt{proc} = \texttt{region} \cdot \texttt{region\_size} + \texttt{local\_rank}$\;
     $\texttt{sizes}[proc] = \texttt{size}$\;
     Non-blocking send of \texttt{buf}[\texttt{region}] to proc\;
  }
  
  MPI\_Allreduce(\texttt{sizes})
  
  \BlankLine%
  \For{$i\gets0$ \KwTo \texttt{sizes}[\texttt{rank}]}{
      Probe for message and receive dynamically\;
  }

  \For {$\texttt{proc} \gets 0$ \KwTo \texttt{region\_size}}{
    \texttt{sizes}[\texttt{proc}] = size of \texttt{buf}[\texttt{proc}]\;
    Non-blocking send of \texttt{buf}[\texttt{proc}] to \texttt{proc}\;
  }

  \BlankLine%
  MPI\_Allreduce(\texttt{sizes})
  \BlankLine%
  
  \For{$i\gets0$ \KwTo \texttt{sizes}[\texttt{rank}]}{
      Probe for message and receive dynamically\;
  }
  
  Wait for all sends to complete\;
    \caption{Locality-Aware Personalized}\label{alg:personalized_loc}
\end{algorithm2e}
\begin{algorithm2e}[ht!]
  \DontPrintSemicolon%
  \KwIn{\texttt{rank} \tcc*{Process ID}
        \texttt{args} \tcc*{\texttt{MPIX\_Alltoall\textcolor{red}{v}\_crs} Arguments}
        \texttt{local\_rank} \tcc*{Local rank of process within region}
        \texttt{region\_size} \tcc*{Number of processes per region}
        }

  \BlankLine%
  \BlankLine%

  \For{$i\gets0$ \KwTo \texttt{send\_nnz}}{
      \texttt{proc} = \texttt{dest}[$i$]\;
      
      \texttt{region} = \texttt{get\_region(proc)}\;
      
      \texttt{size} = \texttt{sendcount\textcolor{red}{s}}\textcolor{red}{[i]}\;
      
      Copy \texttt{proc}, \textcolor{red}{size}, and \texttt{sendvals}[\texttt{ctr}:\texttt{ctr}+\texttt{size}] into \texttt{buf}[\texttt{region}]\;

      $\texttt{ctr} \gets \texttt{ctr} + \texttt{size}$\;
  }

  \ForEach{\texttt{region} to which \texttt{rank} sends}{
     $\texttt{proc} = \texttt{region} \cdot \texttt{region\_size} + \texttt{local\_rank}$
     Non-blocking send of \texttt{buf}[\texttt{region}] to \texttt{proc}\;
  }

  \BlankLine%
  \While{All sends have not completed}{
      \If {Non-blocking probe finds a message}{
          Dynamically receive \texttt{buf} from \texttt{origin\_proc}\;
          \ForEach{\texttt{proc}, \texttt{size}, \texttt{indices} in \texttt{buf}}{
             Copy \texttt{origin\_proc}, \texttt{size}, \texttt{indices} into \texttt{local\_buf}[\texttt{proc}]\;
          }
      }
  }

  Non-blocking barrier\;

  \While{Non-blocking barrier has not completed}{
      \If {Non-blocking probe finds a message}{
          Dynamically receive \texttt{buf} from \texttt{origin\_proc}\;
          \ForEach{\texttt{proc}, \texttt{size}, \texttt{indices} in \texttt{buf}}{
             Copy \texttt{origin\_proc}, \texttt{size}, \texttt{indices} into \texttt{local\_buf}[\texttt{proc}]\;
          }
      }
  }

  \For {$\texttt{proc} \gets 0$ \KwTo \texttt{region\_size}}{
    \texttt{sizes}[\texttt{proc}] = size of \texttt{buf}[\texttt{proc}]\;
    Non-blocking send of \texttt{buf}[\texttt{proc}] to \texttt{proc}\;
  }

  \BlankLine%
  MPI\_Allreduce(\texttt{sizes})
  \BlankLine%
  
  \For{$i\gets0$ \KwTo \texttt{sizes}[\texttt{rank}]}{
      Probe for message and receive dynamically\;
  }
  
  Wait for all sends to complete\;
    \caption{Locality-Aware Nonblocking}\label{alg:n onblocking_loc}\end{algorithm2e}
Both locality-aware optimizations consist of concatenating all messages that a process $p$ must send to any process within a single region.  For example, the node-aware version of this method would aggregate all messages that process $p$ sends to any process on node $n$, and send all data as a single message to a corresponding process.  The corresponding process is determined to be the process in the region of destination with the same local rank as the sending process $p$, where local rank is determined as the rank of $p$ within its region.  For instance, if there are \texttt{PPN} processes per region and ranks are laid out sequentially across the regions, each process $p$ has local rank $\frac{p}{\texttt{PPN}}$.  A process with local rank $r$ communicates only with other processes of local rank $r$ during the inter-region step.

After the inter-region communication completes with either the personalized or non-blocking approach, all data is redistributed within the region.  This is currently implemented as a personalized algorithm due to the fact that there are often a small number of processes within each socket or node, and they typically will redistribute data to a large percentage of other processes within the region.  In recent years, the number of cores per node have largely increased.  On emerging systems, such as sapphire rapids, it is possible that some sparsity patterns would be better optimized with the non-blocking algorithm within each region.  Additional possible optimizations include using an \texttt{MPI\_Alltoallv} for dense intra-region redistribution.  However, as inter-region communication bottlenecks the SDDE, intra-region communication optimizations are beyond the scope of this paper.

\section{Results}~\label{sec:results}
This section analyzes all SDDE algorithms across a subset of matrices from the Suitesparse Matrix Collective~\cite{suitesparse}.  The subset of matrices is sequential in order on the Suitesparse website when ordered by non-zeros and are four matrices with non-zero counts closest to $25$ million.  These matrices are small enough to be stored on just two nodes but also large enough to be scaled out to $64$ nodes.  

The SDDE algorithms are analyzed on the Quartz supercomputer at Lawrence Livermore National Lab.  Each node of Quartz contains two Intel Xeon E5-2695 v4 processors, totalling 36 available CPU cores per node.  All algorithms are tested on both system versions of MPI available on Quartz, OpenMPI 4.1.2 and Mvapich2 2.3.7.  For simplicity, $32$ processes per node are used in each test, allowing for power of $2$ process counts.

The \texttt{MPIX\_Alltoall\_crs} is analyzed across the subset of Suitesparse matrices with OpenMPI in Figure~\ref{fig:alltoall_openmpi} and Mvapich2 in Figure~\ref{fig:alltoall_mvapich}.  For these tests, the sparse dynamic data exchange communicates data sizes based on the sparsity pattern of each matrix, so all messages contain only a single integer.  For instance, if a process $p$ holds non-zeros in columns $x$, $y$, and $z$, for which corresponding rows are stored on process $q$, the SDDE will contain a message with the integer $3$ sent from process $p$ to $q$, indicating that during subsequent operations, $q$ should send $3$ elements to process $p$.  

The costs of the various \texttt{MPIX\_Alltoall\_crs} algorithms are displayed as black lines, while the maximum number of inter-node messages sent by any process are displayed as red dots.  The locality-aware SDDE methods communicate only the aggregated number of inter-node messages while the personalized, non-blocking, and RMA approaches all communicate the standard message count.  

For all matrices, the maximum number of inter-node messages is greatly decreased with aggregation.  At large scales, the locality-aware non-blocking SDDE typically outperforms the other methods due to the lack of reduction along with the low message count.  Furthermore, the scalability of this method is greatly improved over the other approaches, allowing for increased efficiency at even larger scales.  Note, some node counts have no results on OpenMPI due to a UCX error during RMA communication.

Figures~\ref{fig:alltoallv_openmpi} and~\ref{fig:alltoallv_mvapich} analyze the costs of the various \texttt{MPIX\_Alltoallv\_crs} algorithms using OpenMPI and Mvapich2, respectively.  The variable-sized SDDE methods form the communication pattern that is required to perform sparse matrix operations with the given matrices.  As a result, the message sizes vary, with each message containing indices that should be communicated during subsequent sparse matrix operations.  For example, if a process $p$ holds non-zeros in columns $x$, $y$, and $z$, for which corresponding rows are stored on process $q$, these SDDEs will consist of process $p$ sending a message of size $3$ to $q$ containing the numbers $x$, $y$, and $z$.  Standard and aggregated message counts are equivalent to those within the \texttt{MPIX\_Alltoall\_crs} tests.  However, rather than containing only a single integer, each message contains a number of integers equal to the message count, or the number of edges between the two processes.  Note, there are no RMA results as the RMA approach only applies to \texttt{MPIX\_Alltoall\_crs} methods.

The locality-aware non-blocking SDDE method outperforms the others at scale for all matrices other than \texttt{dielFilterV2clx}, which has the smallest message count.  For the other matrices, the locality-aware non-blocking method improves the performance at $64$ nodes by up to 20x.  Furthermore, the scalability of the locality-aware non-blocking algorithm is best for all matrices, with the SDDE algorithm scaling to at least $32$ or $64$ nodes for all matrices.  As a result, this algorithm improves not only performance at large scales, but also allows for sparse matrix operations and the linear solvers that rely on them to be efficiently performed at increasingly large scales.

\section{Conclusions and Future Directions}~\label{sec:conc}
The optimal sparse dynamic data exchange algorithm varies with communication pattern, problem size, process count, and architecture.  As per-region process counts increase, the potential for locality-aware optimizations increases.  The impact of locality-aware aggregation increases with the number of processes per message, and a large number of messages are expected to have a common region of destination.  Parallel applications with large irregular communication constraints are expected to achieve significant benefits from locality-aware SDDE algorithms at strong scaling limits.  When tested across $32$ nodes, with $32$ processes per node, the locality-aware approach achieves up to 20x speedup over the second best approach for matrices with high message counts, while incurring slowdown for matrices that require few messages.  This impact is seen for both the \texttt{MPIX\_Alltoall\_crs} and \texttt{MPIX\_Alltoallv\_crs} methods across both system installations of MPI.  While this paper did not explore locality-aware aggregation for the RMA method, similar concatenation strategies could be used within \texttt{MPI\_Puts} to reduce the synchronization overheads as well as communication costs.

Future performance models are needed to dynamically select the optimal SDDE algorithm for given sparsity patterns on each existing and emerging architecture.  Furthermore, as heterogeneous architectures increase in prevalence, the SDDE algorithm should be adapted for these systems.  There are an increasingly large number of possible paths of data movement on emerging systems, including GPUDirect, copying to a single CPU, and copying portions of data to all available CPU cores, which should be combined with existing algorithms to optimize on emerging heterogeneous architectures.



\bibliographystyle{IEEEtran}
\bibliography{main}

\end{document}